%% file: gft_tsp.tex
\documentclass[10pt,final,journal]{IEEEtran}
\usepackage[utf8]{inputenc}
\usepackage[T1]{fontenc}

\usepackage{cite}

\usepackage{pifont,xcolor}
\newcommand{\cmark}{{\color{black!30!green} \ding{51}}}
\newcommand{\xmark}{\color{red} \footnotesize\ding{55}}

\usepackage[inline]{enumitem}
\newlist{inlinelist}{enumerate*}{1}
\setlist*[inlinelist,1]{label=\roman*),itemjoin={{, }},itemjoin*={{, and }}}

\usepackage{amsmath,amssymb,amsthm,bm}
\usepackage{mathtools}
\usepackage{cancel}
\usepackage{thmtools}
\usepackage{mdframed}
\usepackage{eqnarray}
\DeclareMathOperator{\atan}{atan}
\DeclareMathOperator{\diag}{diag}
\DeclareMathOperator{\dist}{d}
\DeclareMathOperator{\vol}{vol}
\DeclareMathOperator{\grad}{grad}
\DeclareMathOperator{\gqv}{GQV}
\DeclareMathOperator{\gtv}{GTV}
\DeclareMathOperator{\gdv}{GDV}
\DeclareMathOperator{\gqdv}{GQDV}
\DeclareMathOperator{\mse}{MSE}
\DeclareMathOperator{\smoothness}{\Delta}
\DeclareMathOperator{\trace}{tr}
\DeclareMathOperator{\trconj}{H}
\DeclareMathOperator{\CV}{CV}

\theoremstyle{plain}
\newmdtheoremenv{importanttheorem}{Theorem}
\newtheorem{property}{Property}
\newtheorem{definition}{Definition}
\newmdtheoremenv{importantdefinition}[definition]{Definition}
\newtheorem{remark}{Remark}

\usepackage{hhline,tabu}
\usepackage{multirow}
\usepackage{graphicx}
\graphicspath{{figures/}}

\usepackage[labelformat=simple]{subcaption}
\DeclareCaptionLabelSeparator{periodspace}{.\quad}
\usepackage{hyperref}

\def\equationautorefname~#1\null{(#1)\null}

\def\itemautorefname~#1\null{#1\null}
\renewcommand{\Autoref}[1]{%
  \begingroup%
  \renewcommand\itemautorefname{Item}%
  \autoref{#1}%
  \endgroup%
}

\usepackage{etoolbox}
\makeatletter
\patchcmd{\@bibitem}{\ignorespaces}{\label{bib-#1}\ignorespaces}{}{}
\makeatother

\makeatletter
\newcommand\autorefs[1]{\@first@ref#1,@}
\def\@throw@dot#1.#2@{#1}%
\def\@set@refname#1{%
    \edef\@tmp{\getrefbykeydefault{#1}{anchor}{}}%
    \xdef\@tmp{\expandafter\@throw@dot\@tmp.@}%
    \ltx@IfUndefined{\@tmp autorefnameplural}%
         {\def\@refname{\@nameuse{\@tmp autorefname}s}}%
         {\def\@refname{\@nameuse{\@tmp autorefnameplural}}}%
}
\def\@first@ref#1,#2{%
  \ifx#2@\autoref{#1}\let\@nextref\@gobble%
  \else%
    \@set@refname{#1}%
    \@refname~\ref{#1}%
    \let\@nextref\@next@ref%
  \fi%
  \@nextref#2%
}
\def\@next@ref#1,#2{%
   \ifx#2@ and~\ref{#1}\let\@nextref\@gobble%
   \else, \ref{#1}%
   \fi%
   \@nextref#2%
}
\makeatother

\usepackage{xspace}
\def\smoothmat{\ensuremath{\mathbf{M}}\xspace}
\def\smoothmatelem{\ensuremath{M}\xspace}
\def\adja{\ensuremath{\mathbf{A}}\xspace}
\def\adjanorm{\ensuremath{\mathbf{A}^{(\text{norm})}}\xspace}
\def\lapl{\ensuremath{\mathbf{L}}\xspace}
\def\normlapl{\ensuremath{\bm{\mathcal{L}}}\xspace}
\def\rwlapl{\ensuremath{\mathbf{L}_{\text{RW}}}\xspace}

\def\fundmat{\ensuremath{\mathbf{Z}}\xspace}
\def\hilbertmap#1{\ensuremath{\varphi_{#1}}\xspace}

\def\eye{\ensuremath{\mathbf{I}}\xspace}
\def\qmatrix{\ensuremath{\mathbf{Q}}\xspace}
\def\qmatrixelem{\ensuremath{Q}\xspace}
\def\degree{\ensuremath{\mathbf{D}}\xspace}
\def\Ubf{\ensuremath{\mathbf{U}}\xspace}
\def\fourier{\ensuremath{\mathbf{F}}\xspace}
\def\Hbf{\ensuremath{\mathbf{H}}\xspace}
\def\Cbf{\ensuremath{\mathbf{C}}\xspace}

\def\allone{\ensuremath{\mathbf{1}}\xspace}
\def\zerobf{\ensuremath{\mathbf{0}}\xspace}
\def\dbf{\ensuremath{\mathbf{d}}\xspace}
\def\hbf{\ensuremath{\mathbf{h}}\xspace}
\def\nbf{\ensuremath{\mathbf{n}}\xspace}
\def\sbf{\ensuremath{\mathbf{s}}\xspace}
\def\ubf{\ensuremath{\mathbf{u}}\xspace}
\def\ulbf#1{\ensuremath{\mathbf{u_{#1}}}\xspace}
\def\vlbf#1{\ensuremath{\mathbf{v_{#1}}}\xspace}
\def\xbf{{\ensuremath{\bm{x}}}\xspace}
\def\xibf#1{\ensuremath{\bm{x_{#1}}}\xspace}
\def\ybf{\ensuremath{\mathbf{y}}\xspace}
\def\deltaibf#1{\ensuremath{\bm{\delta_{#1}}}\xspace}
\def\epsilonbf{\ensuremath{\bm{\epsilon}}\xspace}

\def\Lambdabf{\ensuremath{\bm{\Lambda}}\xspace}
\def\Sigmabf{\ensuremath{\bm{\Sigma}}\xspace}
 
\newcommand\ncutparamop{\operatorname{-Ncut}}
\def\ncutparam#1#2{\ensuremath{(#1,#2)\ncutparamop}}

\def\textgfm#1#2{$(#1,#2)$-graph Fourier mode}
\def\textgfms#1#2{$(#1,#2)$-graph Fourier modes}
\def\textgft#1#2{$(#1,#2)$-graph Fourier transform}
\def\textgftaccro#1#2{$(#1,#2)$-GFT}

\def\wrt{\textit{w.r.t.}\xspace}
\def\ie{\textit{i.e.}\xspace}
\def\eg{\textit{e.g.}\xspace}

\begin{document}

\setlength{\abovedisplayskip}{4pt}
\setlength{\belowdisplayskip}{4pt}

\title{Irregularity-Aware Graph Fourier Transforms}
\author{Benjamin~Girault,
 Antonio Ortega~\IEEEmembership{Fellow,~IEEE}
    and Shrikanth S.~Narayanan~\IEEEmembership{Fellow,~IEEE}
  \thanks{The authors are with the Signal and Image Processing Institude, University of Southern 
    California, Los Angeles, CA 90089, USA (firstname.lastname@usc.edu).}
  \thanks{Work partially funded by NSF under grants CCF-1410009, CCF-1527874, CCF-1029373.}
  \thanks{The research is based upon work supported by the Office of the Director 
of National Intelligence (ODNI), Intelligence Advanced Research Projects 
Activity (IARPA), via IARPA Contract No 2017- 17042800005. The views and conclusions contained herein are 
those of the authors and should not be interpreted as necessarily 
representing the official policies or endorsements, either expressed or 
implied, of the ODNI, IARPA, or the U.S. Government. The U.S. Government 
is authorized to reproduce and distribute reprints for Governmental 
purposes notwithstanding any copyright annotation thereon.}}

\maketitle

\begin{abstract}
In this paper, we present a novel generalization of 
the graph Fourier transform (GFT). Our approach is based on 
separately considering the definitions
of signal energy and signal variation, leading to several possible
orthonormal GFTs. Our approach includes traditional definitions of
the GFT as special cases, while also leading to new GFT designs that
are better at taking into account the irregular nature of the graph. 
As an illustration, in the context of 
sensor networks we use the Voronoi cell area of vertices in our GFT
definition, showing that it leads to a more sensible definition of
graph signal energy even when sampling is highly irregular.
\end{abstract}


\section{Introduction}

With the ever growing deluge of data, graph signal processing has been proposed for numerous applications in recent years, thanks 
to its ability to study signals lying on irregular discrete structures. Examples include weather 
stations~\cite{Girault.EUSIPCO.2015}, taxi pickups~\cite{SihengChen.GLOBALSIP.2016} or bicycle 
rides~\cite{Hamon.GLOBALSIP.2013,Hamon.THESIS.2015}, people in a social network~\cite{Girault.GRETSI.2013}, or motion capture
data~\cite{Kao.ICIP.2014}. While classical signal processing is typically built on top of regular sampling in time (1D Euclidean 
space), or space (2D Euclidean space), graphs can be applied in irregular domains, as well as for irregular sampling of Euclidean 
spaces~\cite{Belkin.JCSS.2008}. Thus, graph signal processing can be used to process datasets while taking into consideration 
irregular relationships between the data points. 

Successful use of graph signal processing methods for a given application requires identifying: \begin{inlinelist}\item the right 
graph structure\item the right frequency representation of graph signals\end{inlinelist}. The choice of graph structure has been 
studied in recent work on graph learning from 
data~\cite{Daitch.ICML.2009,Dong.TSP.2016,Egilmez.TSP.2017,Pasdeloup.TSIPN.2017,Mei.TSP.2017}. In this paper we focus on the 
second question, namely, given a graph structure, how to extend the classical Fourier transform definition, which relies on a 
regular structure, to a \emph{graph Fourier transform}, which is linked to a discrete irregular structure. 

State of the art methods derive the definition of the graph Fourier transform (GFT) from algebraic representations of the graph 
such as the adjacency matrix \adja, whose entries are the weights of the edges connecting vertices. If $i$ and $j$ are two 
vertices connected through an edge weighted by $w(ij)$, then $A_{ij}=w(ij)$.  In the context of sensor networks, edges are often 
defined by selecting the $K$-nearest neighbors to each vertex, with weights given by a Gaussian kernel of the Euclidean distance. 
This is a setting shown to have good properties in the context of manifold sampling \cite{Hein.JMLR.2007} when the number of 
samples is large. The adjacency matrix is then used to compute the Laplacian matrix $\lapl=\degree-\adja$ with \degree the 
diagonal degree matrix verifying $D_{ii}=d_i=\sum_jw(ij)$. Conventional methods in graph signal processing use the eigenvectors 
of either \adja \cite{Sandryhaila.TSP.2013} or \lapl \cite{Shuman.IEEESP.2013} as graph Fourier modes.

One motivation of this paper comes from physical sensing applications, where data collection is performed by recording on a 
discrete set of points, such as weather readings in a weather station network. In these applications the distribution of the 
sensors is irregular, and so it is not easy to map these measurements back onto a grid in order to use the classical signal 
processing toolbox.
Importantly, in many of these applications, the specific arrangement of sensors is unrelated to the quantity measured. As an 
obvious example, the exact location and distribution of weather stations does not have any influence over the weather patterns in 
the region. Thus, it would be desirable to develop graph signal processing representations that \begin{inlinelist}\item have a 
meaningful interpretation in the context of the sensed data \item are not sensitive to changes in graph structure, and in 
particular its degree of regularity \end{inlinelist}.%
\footnote{Note that studies of stationarity of graph signals, such as \cite{Girault.EUSIPCO.2015}, are concerned with the 
  variability across multiple observations of signals on the {\em same} graph. Instead here we focus on how the choice of 
  different graphs (\ie, different vertex/sensor positions placed in space) affects the spectral representation of the 
  corresponding graph signals (\eg, different discrete samples of the same continuous domain signal).
}

In order to motivate our problem more precisely on this sensor network example, consider a popular definition of 
GFT, based on the eigenvalues of \lapl, with variation for a graph signal \xbf defined as:
\[
  \xbf^{\trconj}\lapl\xbf=\frac{1}{2}\sum_{ij}w(ij)\left|x_i-x_j\right|^2
  \text{.}
\]
A signal $\xbf$ with high variation will have very different values (\ie, large $\left|x_i-x_j\right|^2$) 
at nodes connected by edges with large weights $w(ij)$. This GFT 
definition is such that variation for constant signals is zero, since $\lapl\allone=\mathbf{0}$. This is a nice property, as it
matches definitions of frequency in continuous domain, \ie, a constant signal corresponds to minimal variation and thus frequency 
zero. Note that this property is valid independently of the number and position of sensors in the environment, thus achieving our 
goal of limiting the impact of graph choice on signal analysis. 

In contrast, consider the frequency representation associated to an impulse localized to one vertex in the graph. The impulse 
signal \deltaibf{i} has variation $\deltaibf{i}^{\trconj}\lapl\deltaibf{i}^{\vphantom{\trconj}}=d_i$, where $d_i$ is the degree 
of vertex $i$. Impulses do not have equal variation, so that a highly localized phenomenon in continuous space (or a 
sensor anomaly affecting just one sensor) would produce different spectral signatures depending on the degree of the vertex where 
the measurement is localized. As a consequence, if the set of sensors changes, \eg, because only a subset of sensors are active at 
any given time, the same impulse at the same vertex may have a significantly different spectral representation. 

As an alternative,  
choosing the symmetric normalized Laplacian $\normlapl=\degree^{-1/2}\lapl\degree^{-1/2}$ to define the GFT would lead to  
the opposite result: all impulses would have the same variation, but a constant signal would no longer correspond to the 
lowest frequency of the GFT. Note the importance of graph regularity in this trade-off: the more irregular the degree 
distribution, the more we deviate from desirable behavior for the impulses ($\lapl$) or for the lowest frequency ($\normlapl$).

As further motivation, consider the definition of signal energy, which is typically its $\ell_2$-norm in the literature: 
$E_\xbf=\|\xbf\|_2^2$ \cite{Sandryhaila.TSP.2013,Shuman.IEEESP.2013,Girault.SPL.2015}. Assume that there is an area within the 
region being sensed where the energy of the continuous signal is higher. For this given continuous space signal, energy estimates 
through graph signal energy will depend significantly on the position of the sensors (\eg, these estimates will be higher if 
there are more sensors where the continuous signal energy is higher). Thus, estimated energy will depend on the choice of sensor 
locations, with more significant differences the more irregular the sensor distribution is.

In this paper, we propose a novel approach to address the challenges associated with irregular graph structures by replacing the 
dot product and the $\ell_2$-norm, with a different inner product 
and its induced norm. 
Note that here \emph{irregularity} refers to how irregular the graph structure is, with vertices having local graph 
structures that can vary quite significantly. In particular, a graph can be regular in the sense of all vertices having equal
degree, yet vertices not being homogeneous \wrt to some other respect.
Different 
choices of inner product can be made for different applications. For any of these choices we show that we can compute a set of 
graph signals of increasing variation, and orthonormal with respect to the chosen inner product. These graph signals form the 
basis vectors for novel irregularity-aware graph Fourier transforms that are both theoretically and computationally efficient. 
This framework applies not only to our motivating example of a sensor network, but also to a wide range of applications where 
vertex irregularity needs to be taken into consideration.

In the rest of this paper, we first introduce our main contribution of an irregularity-aware graph Fourier transform using a 
given graph signal inner product and graph signal energy (\autoref{sec:energy}). We then explore the definition of graph filters 
in the context of this novel graph Fourier transform and show that they share many similarities with the classical definition of 
graph filters, but with more richness allowed by the choice of inner product (\autoref{sec:filters}). We then discuss specific choices 
of inner product, including some that correspond to GFTs known in the literature, as well novel some novel ones  
(\autoref{sec:choosing_q}). Finally, we present two applications of these irregularity-aware graph Fourier transforms: vertex 
clustering and analysis of sensor network data (\autoref{sec:experiments}).

\section{Graph Signal Energy: Leaving the Dot Product}
\label{sec:energy}

One of the cornerstones of classical signal processing is the Fourier transform, and one of its essential properties is 
orthogonality. Indeed, this leads to the Generalized Parseval Theorem: The inner product in the time $\langle \xbf,\ybf\rangle$ and spectral 
$\langle \widehat{\xbf},\widehat{\ybf}\rangle$ domains of two signals are equal. 
In this section, we propose a generalization to the definition of the GFT. Our key observation is to note 
that the choice of an inner product is a parameter in the definition of the GFT, and the usual dot product is 
not the only choice available for a sound definition of a GFT verifying Parseval's Theorem.

\subsection{Graph Signal Processing Notations}
\label{sec:energy:notations}

Let $\mathcal{G}=\left(\mathcal{V},\mathcal{E},w\right)$ be a \emph{graph} with $\mathcal{V}=\left\{1,\dots,N\right\}$ its 
\emph{vertex} set,%
\footnote{Although vertices are indexed, the analysis should be independent of the specific indexing chosen, and any other 
indexing shall lead to the same outcome for any vertex for a sound GSP setting.}
$\mathcal{E}\subseteq\mathcal{V}\times\mathcal{V}$ its \emph{edge} set, and $w:\mathcal{E}\rightarrow\mathbb{R}_+$ the 
\emph{weight} function of the edges (as studied in \cite{Girault.THESIS.2015}, these weights should describe the similarity 
between vertices). For simplicity, we will denote the edge from $i$ to $j$ as $ij$. A graph is said \emph{undirected} if for any 
edge $ij$, $ji$ is also an edge, and their weights are equal: $w(ij)=w(ji)$. A \emph{self-loop} is an edge connecting a vertex to 
itself.

A graph is algebraically represented by its \emph{adjacency matrix}%
\footnote{Throughout this paper, bold uppercase characters represent matrices and bold lowercase characters represent 
column vectors.} 
\adja verifying $A_{ij}=0$ if $ij$ is not an edge, and $A_{ij}=w(ij)$ otherwise.%
\footnote{Note that $A_{ij}$ corresponds here to the edge from $i$ to $j$, opposite of the convention used in 
\cite{Sandryhaila.TSP.2013}. This allows for better clarity at the expense of the interpretation of the filtering operation 
$\adja \xbf$. Here, $\left(\adja \xbf\right)_i=\sum_jA_{ij}x_j$ is the diffusion along the edges in reverse direction, from $j$ 
to $i$.}
We use also a normalization of the adjacency matrix by its eigenvalue of largest magnitude $\mu_\text{max}$ such that 
$\adjanorm=\frac{1}{\mu_\text{max}}\adja$  \cite{Sandryhaila.TSP.2014}. We denote $d_i=\sum_j A_{ij}$ the \emph{degree} of vertex 
$i$, and \degree the diagonal \emph{degree matrix} having \dbf as diagonal. When the graph is undirected, the \emph{Laplacian 
matrix} is the matrix $\lapl=\degree-\adja$. Two normalizations of this matrix are frequently used in the literature: the 
\emph{normalized Laplacian matrix} $\normlapl=\degree^{-1/2}\lapl\degree^{-1/2}=\eye-\degree^{-1/2}\adja\degree^{-1/2}$ and the 
\emph{random walk Laplacian} $\rwlapl=\degree^{-1}\lapl=\eye-\degree^{-1}\adja$.

In this graph signal processing literature, the goal is to study signals lying on the vertices of a graph. More precisely, a \emph{graph signal} is 
defined as a function of the vertices $x:\mathcal{V}\rightarrow\mathbb{R}\text{ or }\mathbb{C}$. The vertices being indexed by 
$\left\{1,\dots,N\right\}$, we represent this graph signal using the column vector $\xbf=\left[x_1,\dots,x_N\right]^T$.

Following the literature, the 
GFT is defined as the mapping from a signal \xbf to its spectral 
content $\widehat{\xbf}$ by orthogonal projection onto a set of \emph{graph Fourier modes} (which are themselves graph signals) 
$\left\{\ulbf{0},\dots,\ulbf{N-1}\right\}$ and such that 
$x_i=\sum_l\widehat{x}(l)\left[\ulbf{l}\right]_i=\left[\Ubf\mathbf{\widehat{x}}\right]_i$, where $\Ubf=\left[\ulbf{0}\cdots 
\ulbf{N-1}\right]$. We denote $\fourier^{-1}=\Ubf$ the \emph{inverse GFT matrix} and $\fourier=\Ubf^{-1}$ the \emph{GFT matrix}.

Furthermore, we assume that there is an operator $\smoothness:\mathbb{C}^{\mathcal{V}}\rightarrow\mathbb{R}_+$ quantifying how 
much variation a signal shows on the graph. $\smoothness$ typically depends on the graph chosen. Several examples are given on 
\autorefs{tab:energy:variation_hpsd,tab:energy:variation_non_hpsd}. The graph variation $\smoothness$ is related to a frequency 
as variation increases when frequency increases. The \emph{graph frequency} of the graph Fourier mode \ulbf{l} is then defined as 
$\lambda_l=\smoothness(\ulbf{l})$.%
\footnote{We also used before $\lambda_l=\sqrt{\smoothness(\ulbf{l})}$ when $\smoothness$ is a quadratic variation 
\cite{Girault.SPL.2015}. In this paper, however, the difference between the two does not affect the results.}

In summary, the two ingredients we need to perform graph signal processing are \begin{inlinelist}\item the graph Fourier matrix 
\fourier defining the projection on a basis of elementary signals (\ie, the graph Fourier modes) \item the variation operator 
$\smoothness$ defining the graph frequency of those elementary signals\end{inlinelist}. Our goal in this paper is to propose a 
new method to choose the graph Fourier matrix \fourier, one driven by the choice of graph signal inner product. For clarity and 
conciseness, we focus on undirected graphs, but our definitions extend naturally to directed graphs. An in-depth study of 
directed graphs will be the subject of a future communication.  

\subsection{Norms and Inner Products}
\label{sec:energy:norms}

\begin{table}[t]
  \caption{Hermitian positive semidefinite (HPSD) graph variation operators. Here $\gqv$ is the Graph Quadratic Variation (see \autoref{sec:choosing_q:dot_prod:adja}).}
  \label{tab:energy:variation_hpsd}
  \centering
  \def\arraystretch{1.3}
  \begin{tabu}{|rl|c|}
    \hhline{|==|=|}
    \rowfont[c]{\bfseries}
    \multicolumn{2}{|c|}{Name (\smoothmat)} & $\smoothness(\xbf)$ \\
    \hline
    \cite{Shuman.IEEESP.2013}\hspace*{-0.3cm} & Comb. Lapl. (\lapl) & $\frac{1}{2}\sum_{ij} w(ij) |x_i - x_j|^2$ \\
    \cite{Shuman.IEEESP.2013}\hspace*{-0.3cm} & Norm. Lapl. (\normlapl) & $\frac{1}{2}\sum_{ij} w(ij) 
      \left|\frac{x_i}{\sqrt{d_i}} - \frac{x_j}{\sqrt{\smash[b]{d_j}}}\right|^2$ \\
    \cite{Sandryhaila.TSP.2014}\hspace*{-0.3cm} & $\gqv$ ($(\eye-\adjanorm)^2$) & $\sum_i |x_i - [\adjanorm\xbf]_i|^2$ \\
    \hhline{|==|=|}
  \end{tabu}
\end{table}

In the literature on graph signal processing, it is desirable for the GFT to orthogonal, \ie, \emph{the 
graph Fourier transform is an orthogonal projection on an orthonormal basis of graph Fourier modes}. Up to now, orthogonality has 
been defined using the dot product: $\langle \xbf,\ybf\rangle=\ybf^{\trconj}\xbf$, with $.^{\trconj}$ the transpose conjugate 
operator. We propose here to relax this condition and explore the benefits of choosing an alternative inner product on graph 
signals to define orthogonality.

First, note that $\langle .,.\rangle_\qmatrix$ is an inner product on graph signals, if and only if there exists a 
\emph{Hermitian positive definite} (HPD) matrix \qmatrix such that:
\[
  \langle \xbf,\ybf\rangle_\qmatrix=\ybf^{\trconj}\qmatrix \xbf
  \text{,}
\]
for any graph signals \xbf and \ybf. We refer to this inner product as the \emph{\qmatrix-inner product}. With this notation, the standard 
dot product is the \eye-inner product. Moreover, the \qmatrix-inner product induces the \qmatrix-norm:
\[
  \|\xbf\|_\qmatrix
    =\sqrt{\langle \xbf,\xbf\rangle_\qmatrix}
    =\sqrt{\xbf^{\trconj}\qmatrix \xbf\vphantom{\langle \xbf,\xbf\rangle_\qmatrix}}
  \text{.}
\]
Therefore, an orthonormal set for the \qmatrix-inner product is a set $\{\ulbf{l}\}_l$ verifying:
\[
  \ulbf{k\vphantom{l}}^{\trconj}\qmatrix \ulbf{l}^{\vphantom{\trconj}}=\begin{cases}
                                                    0 & \text{if }k\neq l \\
                                                    1 & \text{otherwise,} \\
                                                 \end{cases}
\]
\ie, $\Ubf^{\trconj}\qmatrix\Ubf=\eye$ with $\Ubf=[\cdots \ulbf{l} \cdots]$.

Although any HPD matrix \qmatrix defines a proper inner product, we will be mostly focusing on diagonal matrices 
$\qmatrix=\diag(q_1,\dots,q_N)$. In that case, the squared \qmatrix-norm of a graph signal \xbf, \ie, its energy, is a weighted 
sum of its squared components:
\[
  \|\xbf\|_\qmatrix^2=\sum_{i\in\mathcal{V}} q_i|x_i|^2
  \text{.}
\]
Such a norm is simple but will be shown to yield interesting results in our motivating example of sensor networks (see 
\autoref{sec:experiments:sensor_nets}). Essentially, if $q_i$ quantifies how important vertex $q_i$ is on the graph, the energy 
above can be used to account for irregularity of the graph structure by correctly balancing vertex importance. Examples of 
diagonal matrix \qmatrix include \eye and \degree, but more involved definitions can be chosen such as with the 
illustrating example of a ring in \autoref{sec:energy:gfm_ring_example} or our novel approach based on 
Voronoi cells in \autoref{sec:choosing_q:voronoi}.

Moreover, endowing the space of graph signals with the \qmatrix-inner product yields the Hilbert space 
$\mathcal{H}_\mathcal{G}(\qmatrix)=(\mathbb{C}^\mathcal{V},\langle .,.\rangle_\qmatrix)$ of graph signals of $\mathcal{G}$. This 
space is important for the next section where we generalize the graph Fourier transform.

\begin{remark}\label{rem:energy:hilbert}
  To simplify the proofs, we observe that many results from matrix algebra that rely on diagonalization in an orthonormal basis 
  \wrt the dot product can actually be extended to the \qmatrix-inner product using the following isometric operator on Hilbert 
  spaces:
  \[
    \begin{array}{l|rcl}
    \hilbertmap{} \colon & \mathcal{H}_\mathcal{G}(\qmatrix) & \longrightarrow & \mathcal{H}_\mathcal{G}(\eye) \\
        & \xbf & \longmapsto & \qmatrix^{1/2}\xbf \end{array}
    \text{.}
  \]
  Since \qmatrix is invertible, \hilbertmap{} is invertible, and an orthonormal basis in one space is mapped to an orthonormal 
  basis in the other space, \ie, $\{\ulbf{l}\}_l$ is an orthonormal basis of $\mathcal{H}_\mathcal{G}(\qmatrix)$ if and only if 
  $\{\hilbertmap{}(\ulbf{l})\}_l$ is an orthonormal basis of $\mathcal{H}_\mathcal{G}(\eye)$.
\end{remark}
  
For example, if $h$ is a linear operator of $\mathcal{H}_\mathcal{G}(\qmatrix)$ such that $h(\xbf)=\Hbf\xbf$, then 
$\widetilde{h}\colon \ybf\mapsto \hilbertmap{}(h(\hilbertmap{}^{-1}(\ybf)))=\qmatrix^{1/2}\Hbf\qmatrix^{-1/2}\ybf$ is an 
operator of $\mathcal{H}_\mathcal{G}(\eye)$. Also, the eigenvectors of \Hbf are orthonormal \wrt the $\qmatrix$-inner product if 
and only if the eigenvectors of $\qmatrix^{1/2}\Hbf\qmatrix^{-1/2}$ are orthonormal \wrt the dot product. This applies for 
example to \emph{$\rwlapl=\degree^{-1}\lapl$} the random walk Laplacian. Since 
$\normlapl=\degree^{1/2}(\degree^{-1}\lapl)\degree^{-1/2}$ has orthonormal eigenvectors \wrt the dot product, then the 
eigenvectors of \emph{\rwlapl} are orthonormal \wrt the \degree-inner product (see \autoref{sec:choosing_q:rwlapl}).

\subsection{Contribution 1: Generalized Graph Fourier Modes}
\label{sec:energy:gfm}

\begin{table}[t]
  \caption{Non-HPSD graph variation operators. Here, $\gtv$ is the Graph Total Variation (see 
    \autoref{sec:choosing_q:dot_prod:adja}), and $\gdv$ is the graph directed variation (see 
    \autoref{sec:energy:gdv_relation}).}
  \label{tab:energy:variation_non_hpsd}
  \centering
  \def\arraystretch{1.3}
  \begin{tabu}{|rl|c|}
    \hhline{|==|=|}
    \rowfont[c]{\bfseries}
    \multicolumn{2}{|c|}{Name} & $\smoothness(\xbf)$ \\
    \hline
    \cite{Sandryhaila.TSP.2014} & $\gtv$ & $\sum_i \left|x_i - [\adjanorm\xbf]_i\right|$ \\
    \cite{Sardellitti.IEEEST.2017} & $\gdv$ & $\sum_{ij} w(ij) [x_i-x_j]_+$ \\
    \hhline{|==|=|}
  \end{tabu}
\end{table}

\begin{importantdefinition}[Generalized Graph Fourier Modes]\label{def:energy:gfm}
  Given the Hilbert space $\mathcal{H}_\mathcal{G}(\qmatrix)$ of graph signals and the variation operator $\smoothness$, the set of 
  \emph{\textgfms{\smoothness}{\qmatrix}} is defined as an orthonormal basis of graph signals $\{\ulbf{l}\}_l$ solution to the 
  following sequence of minimization problems, for increasing $L\in\{0,\dots,N-1\}$:
  \begin{equation}\label{eq:energy:gfm_min}
    \min_{\ulbf{L}}\smoothness(\ulbf{L})
    \qquad\text{subj. to}\quad
    \Ubf_{\mathbf{L}}^{\trconj}\qmatrix \Ubf_{\mathbf{L}}=\eye
    \text{,}
  \end{equation}
  where $\Ubf_{\mathbf{L}}=[\ulbf{0}\dots \ulbf{L}]$.
\end{importantdefinition}

\input{ring_graph}

This definition is mathematically sound since the search space of each minimization problem is non empty. Note also that this 
definition relies on only two assumptions: \begin{inlinelist}\item \qmatrix is an inner product matrix \item $\smoothness$ maps 
graph signals to real values\end{inlinelist}.%
\footnote{\qmatrix may depend on $\smoothness$, but no such relation is required in \autoref{def:energy:gfm}. An example is given 
at the end of this section where $q_i=\smoothness(\deltaibf{i})$ is desirable.}

Among the examples of graph variation examples given in \autorefs{tab:energy:variation_hpsd,tab:energy:variation_non_hpsd}, those 
of \autoref{tab:energy:variation_hpsd} share the important property of being quadratic forms:

\begin{definition}[Hermitian Positive Semi-Definite Form]
  The variation operator $\smoothness$ is a \emph{Hermitian positive semi-definite (HPSD) form} if and only if 
  $\smoothness(\xbf)=\xbf^{\trconj}\smoothmat\xbf$ and \smoothmat is a Hermitian positive semi-definite matrix.
\end{definition}

When $\smoothness$ is an HPSD form it is then algebraically characterized by the matrix \smoothmat. In what follows 
we denote the graph variation operator as \smoothmat whenever $\smoothness(\xbf)=\xbf^{\trconj}\smoothmat\xbf$ is 
verified. Examples from the literature of graph variation operators that are HPSD forms are shown on 
\autoref{tab:energy:variation_hpsd}, and non-HPSD ones are shown on \autoref{tab:energy:variation_non_hpsd}.

The following theorem is of utmost importance as it relates the solution of \autoref{eq:energy:gfm_min} to a generalized 
eigenvalue problem when $\smoothness$ is an HPSD form:

\begin{importanttheorem}\label{thm:energy:gfm_hermitian}
  If $\smoothness$ is an HPSD form with HPSD matrix \smoothmat,
  then $\{\ulbf{l}\}_l$ is a set of \textgfms{\smoothness}{\qmatrix} if and only if the graph signals $\{\ulbf{l}\}_l$ solve the 
  following generalized eigenvalue problems for increasing eigenvalues $\lambda_l=\smoothness(\ulbf{l})$:
  \[
    \smoothmat\ulbf{l}=\lambda_l\qmatrix\ulbf{l}
    \text{,}
  \]
  with $\|\ulbf{l}\|_\qmatrix^2=1$.
\end{importanttheorem}

\begin{proof}
  Observing that $\xbf^{\trconj}\smoothmat \xbf=\langle \qmatrix^{-1}\smoothmat \xbf,\xbf\rangle_\qmatrix$, we can show that 
  \smoothmat Hermitian is equivalent to $\qmatrix^{-1}\smoothmat$ being a self-adjoint operator on the Hilbert space 
  $\mathcal{H}_\mathcal{G}(\qmatrix)$. Therefore, using the spectral theorem, there exists an orthonormal basis $\{\ulbf{l}\}_l$ 
  of $\mathcal{H}_\mathcal{G}(\qmatrix)$ of eigenvectors of $\qmatrix^{-1}\smoothmat$, such that:
  \[
    \smoothness(\ulbf{l})=\ulbf{l}^{\trconj}\smoothmat \ulbf{l}
      =\ulbf{l}^{\trconj}\qmatrix\qmatrix^{-1}\smoothmat \ulbf{l}
      =\lambda_l\ulbf{l}^{\trconj}\qmatrix \ulbf{l}=\lambda_l
    \text{.}\qedhere
  \]
\end{proof}

The relation above between the variation of the graph Fourier modes and the eigenvalues of \smoothmat also shows that, just as 
approaches of the literature based on the Laplacian \cite{Shuman.IEEESP.2013} or on the adjacency matrix 
\cite{Sandryhaila.TSP.2013}, the set of graph Fourier modes is not unique when there are multiple eigenvalues: If 
$\smoothness(\ulbf{l})=\smoothness(\ulbf{k})$, then if $\vlbf{l}=\alpha \ulbf{l}+(1-\alpha)\ulbf{k}$ and 
$\vlbf{k}=(1-\alpha)\ulbf{l}+\alpha \ulbf{k}$, we have $\smoothness(\ulbf{l})=\smoothness(\vlbf{l})=\smoothness(\vlbf{k})$ and 
$\langle \vlbf{l},\vlbf{k}\rangle_\qmatrix=0$, for any $\alpha\in[0,1]$. This remark will be important in \autoref{sec:filters},  
as it is desirable for the processing of graph signals to be independent of the particular 
choice of \textgfms{\smoothness}{\qmatrix}.

\subsection{Example with the Ring Graph: Influence of \texorpdfstring{\qmatrix}{Q}}
\label{sec:energy:gfm_ring_example}

To give intuitions on the influence of \qmatrix on the \textgfms{\smoothness}{\qmatrix}, we study the example of a ring graph 
with the Laplacian matrix as graph variation operator matrix: $\smoothmat=\lapl$. In \autoref{fig:ring_graph}, we show the 
\textgfms{\lapl}{\qmatrix} of a ring with $8$ vertices for three choices of \qmatrix:
\begin{itemize}
  \item $\qmatrix=\diag(0.1,1,\dots,1)$ with a less important vertex 1,
  \item $\qmatrix=\eye$, \ie, the combinatorial Laplacian GFT,
  \item $\qmatrix=\diag(10,1,\dots,1)$ with a more important vertex 1.
\end{itemize}
Several observations can be made. First of all, any \textgfm{\lapl}{\eye} which is zero on vertex 1
(\ie, $\ubf_1$, $\ubf_3$, $\ubf_5$) is also an \textgfm{\lapl}{\qmatrix} with the same graph variation. Intuitively, this vertex 
having zero value means that it has no influence on the mode (here, $\qmatrix\ulbf{l}=\ulbf{l}$), hence its importance does not 
affect the mode.

For the remaining modes, we make several observations. First of all, we consider the spectral representation of a highly 
localized signal on vertex 1, such as \deltaibf{1} pictured in the last column of \autoref{fig:ring_graph}. While 
$\qmatrix=\eye$ involves many spectral components of roughly the same power, the other two cases $q_1=0.1$ and $q_1=10$ are 
distinctive. Indeed, in the first case, we have $\widehat{\deltaibf{1}}(7)$ that is large (almost an order of magnitude larger),
interpreted by a graph Fourier mode $[\ulbf{7}]$ that is close to our localized signal. On the other hand, for $q_1=10$ the two
largest Fourier components of \deltaibf{1} are $\widehat{\deltaibf{1}}(0)$ and $\widehat{\deltaibf{1}}(1)$ with $[\ulbf{1}]$ 
being close to our localized signal. In other words, $q_i$ shifted the impulse \deltaibf{1} to the higher spectrum when small
(unimportant vertex $i$) or the lower spectrum when large (important vertex $i$).

These cases give intuitions on the impact of \qmatrix, and ultimately on how to choose it.

\subsection{Discussion on \smoothmat}
\label{sec:energy:gfm_discussion}

When $\smoothness$ is and HPSD from, we can rewrite the minimization of \autoref{eq:energy:gfm_min} as a generalized Rayleigh 
quotient minimization. Indeed, the $L^\text{th}$ minimization problem is also given by:
\[
  \min_{\xbf:\Ubf_{\mathbf{L}}^{\trconj}\qmatrix\xbf=\zerobf} \smoothness\left(\frac{\xbf}{\|\xbf\|_\qmatrix}\right)
  \text{,}
\]
which is exactly a Rayleigh quotient minimization when $\smoothness$ is an HPSD form of HPSD matrix \smoothmat since:
\[
  \smoothness\left(\frac{\xbf}{\|\xbf\|_\qmatrix}\right)=\frac{\xbf^{\trconj}\smoothmat\xbf}{\xbf^{\trconj}\qmatrix\xbf}
  \text{.}
\]
For example, this allows the study of bandpass signals using spectral proxies as in \cite{Anis.TSP.2016}.

Having an HPSD form for $\smoothness$ also has two advantages, the first one being a simple solution to 
\autoref{eq:energy:gfm_min} as stated in \autoref{thm:energy:gfm_hermitian}. But more importantly, this solution is efficient%
\footnote{Efficiency comes here from the cost of computing generalized eigenvalues and eigenvectors compared to the when 
\smoothmat is not HPSD or \qmatrix is not HPD.}
to compute through the generalized eigenvalue problem since both \qmatrix and \smoothmat are Hermitian \cite[Section 
8.7]{Golub.BOOK.1996}. 
Therefore, instead of inverting \qmatrix and computing eigenvalues and eigenvectors of the non-Hermitian matrix 
$\qmatrix^{-1}\smoothmat$, the two matrices \smoothmat and \qmatrix can be directly used to compute the 
\textgfms{\smoothmat}{\qmatrix} using their Hermitian properties.

In some applications, it may also be desirable to have \qmatrix depend on \smoothmat. In our motivating example the key 
variations of the constant signal \allone and of the impulses \deltaibf{i} are given by:
\begin{align*}
  \smoothness(\allone) &= \sum_{ij}\smoothmatelem_{ij} &
  \smoothness(\deltaibf{i}) &= \smoothmatelem_{ii}
  \text{.}
\end{align*}
In the case of \allone, variation should be zero, hence $\sum_{ij}\smoothmatelem_{ij}=0$. However, for \deltaibf{i}, the 
variations above cannot be directly compared from one impulse to another as those impulses have different energy. Variations of 
the energy normalized signals yields:
\[
  \smoothness\left(\frac{\deltaibf{i}}{\|\deltaibf{i}\|_\qmatrix}\right) = \frac{\smoothmatelem_{ii}}{\qmatrixelem_{ii}}
  \text{.}
\]
We advocated for a constant (energy normalized) variation. In effect, this leads to a relation between \qmatrix and \smoothmat 
given by constant $\smoothmatelem_{ii}/\qmatrixelem_{ii}$, \ie, the diagonals of \qmatrix and \smoothmat should be equal, up to a 
multiplicative constant. For instance, choosing $\smoothmat=\lapl$, which leads to $\smoothness(\allone)=0$, and 
$\smoothness(\deltaibf{i})=d_i$, our requirement on \qmatrix leads to $q_i=d_i$, hence $\qmatrix=\degree$. This is the random 
walk Laplacian approach described in \autoref{sec:choosing_q:rwlapl}.

\subsection{Relation to \texorpdfstring{\cite{Sardellitti.IEEEST.2017}}{[\ref{bib-Sardellitti.IEEEST.2017}]}}
\label{sec:energy:gdv_relation}

Finally, we look at the recent advances in defining the graph Fourier transform from the literature, and in particular to 
\cite{Sardellitti.IEEEST.2017} which aims at defining an orthonormal set of graph Fourier modes for a directed graph with 
non-negative weights. After defining the \emph{graph directed variation} as:
\[
  \gdv(\xbf) := \sum_{ij}w(ij)\left[x_i-x_j\right]_+
  \text{,}
\]
where $[x]_+=\max(0,x)$, the graph Fourier modes are computed as a set of orthonormal vectors \wrt the dot product that solve:
\begin{align}\label{eq:energy:gfm_gdv_min_sum}
  \min_{\{\ulbf{l}\}_l} &\sum_l\gdv(\ulbf{l}) &
  \text{subj. to}\quad \Ubf^{\trconj}\Ubf &= \eye
  \text{.}
\end{align}
The straightforward generalization of this optimization problem to any graph variation operator $\smoothness$ and \qmatrix-inner 
product, is then:
\begin{align}\label{eq:energy:gfm_min_sum}
  \min_{\{\ulbf{l}\}_l} &\sum_l\smoothness(\ulbf{l}) &
  \text{subj. to}\quad \Ubf^{\trconj}\qmatrix\Ubf &= \eye
  \text{.}
\end{align}
Unfortunately, we cannot use this generalization when $\smoothness$ is an HPSD form. Indeed, in that case, the sum in 
\autoref{eq:energy:gfm_min_sum} is exactly $\trace(\Ubf^{\trconj}\smoothmat\Ubf)=\trace(\qmatrix^{-1}\smoothmat)$ for any matrix 
\Ubf verifying $\Ubf^{\trconj}\qmatrix\Ubf = \eye$. For example, $\Ubf=\qmatrix^{-1/2}$ solves \autoref{eq:energy:gfm_min_sum}, 
but under the assumption that \qmatrix is diagonal, this leads to a trivial diagonal GFT, modes localized on vertices of the graph.

Note that, for any graph variation, a solution to our proposed minimization problem in \autoref{eq:energy:gfm_min} is also a solution 
to the generalization in \autoref{eq:energy:gfm_min_sum}:

\begin{property}\label{prop:energy:gfm_min_sum_solution}
  If $\{\ulbf{l}\}_l$ is a set of \textgfms{\smoothness}{\qmatrix}, then it is a solution to:
  \[
    \min_{\{\ulbf{l}\}_l}\sum_l\smoothness(\ulbf{L})
    \qquad\text{subj. to}\quad
    \Ubf^{\trconj}\qmatrix \Ubf=\eye
    \text{,}
  \]
  with $\Ubf=[\ulbf{0}\dots \ulbf{N-1}]$.
\end{property}

For example, the set of \textgfms{\gdv}{\eye} is a solution to \autoref{eq:energy:gfm_gdv_min_sum}, hence a set of graph Fourier 
modes according to \cite{Sardellitti.IEEEST.2017}. This property is important as it allows, when direct computation through a 
closed form solution of the graph Fourier modes is not possible, to approximate the \textgfms{\smoothness}{\qmatrix} by first 
using the techniques of \cite{Sardellitti.IEEEST.2017} to obtain the 
\textgfms{\qmatrix^{1/2}\smoothness\qmatrix^{-1/2}}{\eye} and then using \autoref{rem:energy:hilbert}.

Finally, another recent work uses a related optimization function to obtain graph Fourier modes\cite{Shafipour.ARXIV.2017}:
\[
  \min_{\{\ulbf{l}\}_l} \sum_l \left(f_l - \gqdv(\ulbf{l})\right)^2
    \qquad\text{subj. to}\quad
    \Ubf^{\trconj}\Ubf=\eye
  \text{,}
\]
with $f_l=\frac{l - 1}{N - 1}f_\text{max}$ and the \emph{graph quadratic directed variation} defined as:
\[
  \gqdv(\xbf) := \sum_{ij} w(ji)\left[x_i-x_j\right]_+^2
  \text{.}
\]
Beyond the use of a squared directed difference $\left[x_i-x_j\right]_+^2$, the goal of this optimization is to obtain evenly 
distributed graph frequencies, and not orthogonal graph Fourier modes of minimally increasing variation of 
\cite{Sardellitti.IEEEST.2017} and of our contribution. Note that this alternative approach can implement the constraint 
$\Ubf^{\trconj}\qmatrix\Ubf=\eye$ to use an alternative definition of graph signal energy. This is however out of the scope of 
this paper.

\subsection{Contribution 2: Generalized GFT}
\label{sec:energy:gft}

Given the \textgfms{\smoothness}{\qmatrix} in the previous section, and the definition of the inverse graph Fourier transform 
$\fourier^{-1}=\Ubf$ found in the literature and recalled in \autoref{sec:energy:notations}, we can now define the generalized 
graph Fourier transform. Note that $\smoothness$ is not assumed to be an HPSD form anymore.

\begin{importantdefinition}[Generalized Graph Fourier Transform]\label{def:energy:gft}
  Let \Ubf be the matrix of \textgfms{\smoothness}{\qmatrix}. 
  The \emph{\textgft{\smoothness}{\qmatrix} (\textgftaccro{\smoothness}{\qmatrix})} is then:
  \[
    \fourier = \Ubf^{\trconj}\qmatrix
    \text{,}
  \]
  and its inverse is:
  \[
    \fourier^{-1} = \Ubf
    \text{.}
  \]
\end{importantdefinition}

The inverse in \autoref{def:energy:gft} above is a proper inverse since $\fourier\fourier^{-1}=\Ubf^{\trconj}\qmatrix\Ubf=\eye$. 
In \autoref{sec:energy:notations}, we introduced the graph Fourier transform as an orthogonal projection on the graph Fourier 
modes. \Autoref{def:energy:gft} is indeed such a projection since:
\[
  \widehat{x}(l)
    = \left[\fourier \xbf\right]_l
    = \left[\Ubf^{\trconj}\qmatrix \xbf\right]_l
    = \ulbf{l}^{\trconj}\qmatrix \xbf
    = \langle \xbf,\ulbf{l}\rangle_\qmatrix
  \text{.}
\]
Another important remark on \autoref{def:energy:gft} concerns the complexity of computing the graph Fourier matrix. Indeed, this 
matrix is the inverse of the inverse graph Fourier matrix $\fourier^{-1}=\Ubf$ which is directly obtained using the graph Fourier 
modes. Computation of a matrix inverse is in general costly and subject to approximation errors. However, just as classical GFTs 
for undirected graphs, we can obtain this inverse without preforming an actual matrix inverse. Indeed, our GFT matrix is obtained 
through a simple matrix multiplication $\Ubf^{\trconj}\qmatrix=\Ubf^{-1}$ that uses the orthonormal property of the graph Fourier 
modes basis. Additionally, when \qmatrix is diagonal, this matrix multiplication is extremely simple and easy to compute.

One property that is essential in the context of an orthonormal set of graph Fourier modes is Parseval's Identity where the 
energy in the vertex and spectral domains are equal:
\begin{property}[Generalized Parseval's Theorem]\label{prop:energy:parseval}
  The \textgftaccro{\smoothness}{\qmatrix} is an isometric operator from 
  $\mathcal{H}_\mathcal{G}(\qmatrix)$ to $\mathcal{H}_{\smash[t]{\widehat{\mathcal{G}}}}(\eye)$. Mathematically:
  \[
    \left\langle\xbf,\ybf\right\rangle_\qmatrix
      =\left\langle\widehat{\xbf},\widehat{\ybf}\right\rangle_\eye
    \text{.}
  \]
\end{property}
\begin{proof}
  $
    \left\langle\xbf,\ybf\right\rangle_\qmatrix
    =\ybf^{\trconj}\qmatrix \xbf
    =\widehat{\ybf}^{\trconj}\left(\fourier^{-1}\right)^{\trconj}\qmatrix \fourier^{-1} \widehat{\xbf}
    =\widehat{\ybf}^{\trconj}\widehat{\xbf}
    \text{.}
  $
\end{proof}

Finally, \autoref{prop:energy:parseval} bears similarities with \autoref{rem:energy:hilbert}. Indeed, in both cases, there is an 
isometric map from the Hilbert space $\mathcal{H}_\mathcal{G}(\qmatrix)$ to either $\mathcal{H}_\mathcal{G}(\eye)$ or 
$\mathcal{H}_{\smash[t]{\widehat{\mathcal{G}}}}(\eye)$. However, in \autoref{prop:energy:parseval}, the graph signal \xbf is 
mapped to the spectral components $\fourier\xbf$ of \xbf instead of $\qmatrix^{1/2}\xbf$ such that both cases are distinct. 
Intuitively, \hilbertmap{} and $\qmatrix^{1/2}$ reshape the space of graph signals to account for irregularity, while the graph 
Fourier matrix \fourier decomposes the graph signal into elementary graph Fourier modes of distinct graph variation.

\section{Graph Filters}
\label{sec:filters}

In this section we explore the concept of operator on graph signals, and more specifically, operators whose definition is 
intertwined with the graph Fourier transform. Such a relation enforces a spectral interpretation of the operator and ultimately 
ensures that the graph structure plays an important role in the output of the operator.

\subsection{Fundamental Matrix of the GFT}
\label{sec:filters:fundmat}

Before defining graph filters, we need to introduce what we call the \emph{fundamental matrix of the 
\textgftaccro{\smoothness}{\qmatrix}} given its graph Fourier matrix \fourier and the diagonal matrix of graph frequencies 
\Lambdabf:
\[
  \fundmat:=\fourier^{-1}\Lambdabf\fourier=\Ubf\Lambdabf\Ubf^{\trconj}\qmatrix
  \text{.}
\]
Although $\smoothness$ does not appear in the definition above, \fundmat does depend on $\smoothness$ through the graph Fourier 
matrix \fourier. Some authors use the term \emph{shift} for this matrix when $\qmatrix=\eye$. However, the literature uses very 
often the Laplacian matrix \cite{Shuman.IEEESP.2013} as the fundamental matrix, and as a close equivalent to a second order 
differential operator \cite{Hein.JMLR.2007}, it does not qualify as the equivalent of a shift operator. Therefore, we choose the 
more neutral term of \emph{fundamental matrix}. Yet, our definition is a generalization of the literature where such a matrix is 
always diagonalizable in the graph Fourier basis, with graph frequencies as eigenvalues. \autoref{tab:choosing_q:gft} shows 
several classical choices of fundamental matrices depending on \qmatrix and $\smoothness$.

Further assuming that $\smoothness$ is an HPSD form, we have $\fundmat=\qmatrix^{-1}\smoothmat$. As we will see in 
\autoref{sec:choosing_q}, this is consistent with the graph signal processing literature. Noticeably, this matrix is also 
uniquely defined under these conditions, even though the graph Fourier matrix may not be. Moreover, complexity of computing this 
matrix is negligible when \qmatrix is diagonal. A diagonal matrix \qmatrix also leads to \fundmat having the same sparsity as 
\smoothmat since \smoothmat and \fundmat differ only for non-zero elements of \smoothmat. Algorithms whose efficiency is driven 
by the sparsity of \smoothmat are generalized without loss of efficiency to $\qmatrix^{-1}\smoothmat=\fundmat$. This will be the 
case for the examples of this paper.

\subsection{Definitions of Graph Filters}
\label{sec:filters:defs}

We recall here the classical definitions of graph filters and describe how we straightforwardly generalize them using the 
fundamental matrix \fundmat of the GFT.

First of all, given a graph filter \Hbf, we denote by $\widehat{\Hbf}=\fourier\Hbf\fourier^{-1}$ the same graph filter in the 
spectral domain such that $\widehat{\Hbf\xbf}=\widehat{\Hbf}\widehat{\xbf}$. This notation allows to properly study the spectral 
response of a given graph filter. We can now state the three definitions of graph filters.

The first one directly extends the invariance through time shifting. This is the approach followed by \cite{Sandryhaila.TSP.2013}, 
replacing the adjacency matrix by the fundamental matrix of the graph:

\begin{definition}[Graph Filters by Invariance]\label{def:filters:filter_invariance}
  \Hbf is a \emph{graph filter} if and only if it commutes with the fundamental matrix \fundmat of the GFT:
  \[
    \Hbf\fundmat=\fundmat\Hbf
    \text{.}
  \]
\end{definition}

The second definition extends the convolution theorem for temporal signals, where convolution in the time domain is equivalent to 
pointwise multiplication in the spectral domain \cite[Sec. 3.2]{Girault.THESIS.2015}%
\footnote{In \cite{Shuman.ACHA.2016}, the authors also define convolutions as pointwise multiplications in the spectral domain. 
However, the notation used for the GFT is unclear since $\widehat{x}(\lambda_l)$ (instead of $\widehat{x}(l)$) can be 
interpreted as the graph signal \xbf having equal spectral components for equal graph frequencies. Such a requirement actually 
falls within the setting of \autoref{def:filters:filter_polynomial}.}.
The following definition is identical to those in the literature, simply replacing existing choices of GFT with one of our proposed GFTs:

\begin{definition}[Graph Filters by Convolution Theorem]\label{def:filters:filter_conv}
  \Hbf is a \emph{graph filter} if and only if there exists a graph signal $\hbf$ such that:
  \[
    \widehat{\Hbf \xbf}(l) = \widehat{h}(l)\widehat{x}(l)
    \text{.}
  \]
\end{definition}

The third definition is also a consequence of the convolution theorem, but with a different interpretation. Indeed, in the time 
domain, the Fourier transform $\widehat{s}$ of a signal $s$ is a function of the (signed) frequency. Given that there is a finite 
number of graph frequencies, any function of the graph frequency can be written as a polynomial of the graph frequency (through 
polynomial interpolation). We obtain the following definition using the fundamental matrix instead 
\cite{Hammond.ACHA.2011,Shuman.ACHA.2016}:

\begin{definition}[Graph Filters by Polynomials]\label{def:filters:filter_polynomial}
  \Hbf is a \emph{graph filter} if and only if it is a polynomial in the fundamental matrix \fundmat of the GFT:
  \[
    \Hbf=\sum_k h_k\fundmat^k
    \text{.}
  \]
\end{definition}

Interestingly, these definitions are equivalent 
for a majority of graphs where no two graph frequencies are equal. However, in the converse case of a graph with two equal 
graph frequencies $\lambda_k=\lambda_l$, these definitions differ. Indeed, \autoref{def:filters:filter_polynomial} implies that 
$\widehat{h}(k)=\widehat{h}(l)$, while $\widehat{h}(k)$ and $\widehat{h}(l)$ are arbitrary according to 
\autoref{def:filters:filter_conv}. Also, a graph filter according to \autoref{def:filters:filter_invariance} does not necessarily 
verify $\widehat{\Hbf}$ diagonal, since $\widehat{\Hbf u_l} =H_{k,l}u_k+H_{l,l}u_l$ is a valid graph filter, even with 
$H_{k,l}\neq 0$. Choosing one definition over another is an application-dependent choice driven by the meaning of two graph 
Fourier modes of equal graph frequencies. For example, if these modes are highly related, then 
\autoref{def:filters:filter_polynomial} is a natural choice with equal spectral response of the filter, whereas in the opposite 
case of unrelated modes, \autoref{def:filters:filter_conv} allows to account for more flexibility in the design of the filter.

\subsection{Mean Square Error}
\label{sec:filters:mse}

\emph{Mean Square Error} (MSE) is classically used to study the error made by a filter when attempting to recover a signal from a 
noisy input. More precisely, given an observation $\ybf=\xbf+\nbf$ of a signal $\xbf$ with additive random noise $\nbf$ (with 
zero mean), MSE is defined as:
\begin{multline}\label{eq:filters:mse:l2}
  \mse_\Hbf(\xbf):=\mathbb{E}\left[\|\Hbf\ybf - \xbf\|_2^2\right] \\
    =\sum_i \mathbb{E}\left[\left(\left[\Hbf\ybf\right]_i - x_i\right)^2\right]
  \text{.}
\end{multline}
In other words, this is the mean energy of the error made by the filter. However, the definition of energy used above is the dot 
product. As stated before, this energy does not account for irregularity of the structure.

In the general case of an irregular graph, we defined in \autoref{sec:energy:norms} graph signal energy using the \qmatrix-norm. 
We define in this section the energy of an error by the \qmatrix-norm of that error, thus generalizing mean square error into 
the \qmatrix-MSE:
\[
  \mse_\Hbf^{(\qmatrix)}(\xbf):=\mathbb{E}\left[\left\|\Hbf\ybf - \xbf\right\|_\qmatrix^2\right]
  \text{.}
\]
Using the zero mean assumption, this yields the classical bias/variance formula:
\begin{equation}\label{eq:filters:mse:bias_variance}
  \mse_\Hbf^{(\qmatrix)}(\xbf):=
    \underbrace{
      \left\|\Hbf\xbf - \xbf\right\|_\qmatrix^2
      \vphantom{\mathbb{E}\left[\left\|\Hbf\nbf\right\|_\qmatrix^2\right]}
    }_{\text{bias term}}
    +
    \underbrace{\mathbb{E}\left[\left\|\Hbf\nbf\right\|_\qmatrix^2\right]}_{\text{noise variance term}}
\end{equation}
The definition of graph Fourier transform introduced in \autoref{def:energy:gft} is then a natural fit to study this MSE thanks 
to Parseval's identity (\autoref{prop:energy:parseval}):
\[
  \mse_\Hbf^{(\qmatrix)}(\xbf)=
    \left\|\left(\widehat{\Hbf}-\eye\right)\widehat{\xbf}\right\|_2^2
    +
    \mathbb{E}\left[\left\|\widehat{\Hbf}\widehat{\nbf}\right\|_2^2\right]
  \text{.}
\]
Since $\widehat{\Hbf}$ is diagonal (or block diagonal if using \autoref{def:filters:filter_invariance}), studying the bias and 
the variance from a spectral point of view is much simpler. Recalling the meaning of the graph Fourier modes in terms of graph 
variation, this also allows to quantify the bias and variance for slowly varying components of the signal (lower spectrum) to 
quickly varying components (higher spectrum) giving an intuitive interpretation to MSE and the bias/variance trade off across 
graph frequencies.

To have a better intuition on how using the \qmatrix-norm allows to account for irreguarity, we now assume that
\qmatrix is diagonal. Let $\epsilonbf=\Hbf\nbf$ be the filtered noise. The noise variance term above becomes:
\[
  \mathbb{E}\left[\left\|\epsilonbf\right\|_\qmatrix^2\right]=
    \sum_i q_i \mathbb{E}\left[|\epsilon_i|^2\right]
\]
In other words, those vertices with higher $q_i$ have a higher impact on the overall noise energy. Using 
$\mathbb{E}\left[\|\epsilonbf\|_\qmatrix^2\right]=\trace\left(\epsilonbf\epsilonbf^{\trconj}\qmatrix\right)$, we also have:
\[
  \mathbb{E}\left[\|\Hbf\nbf\|_\qmatrix^2\right]
    =\trace\left(\widehat{\Hbf}^{\trconj}\widehat{\Hbf}\fourier\Sigma_\nbf\qmatrix\fourier^{-1}\right)
  \text{,}
\]
where $\Sigmabf_\nbf=\mathbb{E}\left[\nbf\nbf^{\trconj}\right]$ is the covariance matrix of the noise. We introduce a definition 
of \qmatrix-white noise with a tractable power spectrum:

\begin{importantdefinition}[\qmatrix-White Noise]
  The graph signal noise \nbf is said \emph{\qmatrix-White Noise} (\qmatrix-WN) if and only if it is centered 
  $\mathbb{E}\left[\nbf\right]=\zerobf$ and its covariance matrix verifies $\Sigmabf_\nbf=\sigma^2\qmatrix^{-1}$,
  for some $\sigma\geq0$.
\end{importantdefinition}

There are two important observations to make on this definition. In the vertex domain, if we assume a diagonal \qmatrix, this 
noise has higher power on vertices with smaller $q_i$: $\mathbb{E}\bigl[\left|n_i\right|^2\bigr]=\frac{\sigma^2}{q_i}$. 
\emph{Assuming \qmatrix-WN is equivalent to assuming less noise on vertices with higher $q_i$.} Therefore, this definition of 
noise can account for the irregularity of the graph structure through \qmatrix.

Second, the importance of this definition is best seen in the spectral domain. Indeed, the spectral covariance matrix verifies:
\[
  \Sigmabf_{\widehat{\nbf}}=\fourier\Sigma_\nbf\fourier^{\trconj}=\sigma^2\eye
  \text{.}
\]
In other words, a \qmatrix-WN has a flat power spectrum. Note that this is true independently of the variation operator 
$\smoothness$ chosen to define the \textgftaccro{\smoothness}{\qmatrix} matrix \fourier, hence the name of \qmatrix-WN (and not 
$(\smoothness,\qmatrix)$-WN).

The noise variance term in \autoref{eq:filters:mse:bias_variance} under the assumption of a \qmatrix-WN \nbf becomes:
\[
  \mathbb{E}\left[\left\|\Hbf\nbf\right\|_\qmatrix^2\right]
    =\sigma^2\trace\left(\widehat{\Hbf}^{\trconj}\widehat{\Hbf}\right)
  \text{.}
\]
In other words, it is completely characterized by the spectral response of the filter \Hbf. Note that using 
\autoref{def:filters:filter_invariance}, $\widehat{\Hbf}$ is not necessarily Hermitian. Using \autoref{def:filters:filter_conv} 
or \autoref{def:filters:filter_polynomial}, $\widehat{\Hbf}$ is diagonal and the noise variance term simplifies to:
\[
  \mathbb{E}\left[\left\|\Hbf\nbf\right\|_\qmatrix^2\right]
    =\sigma^2\sum_l\left|\widehat{h}(l)\right|^2
  \text{.}
\]

\section{Choosing the Graph Signal Energy Matrix \qmatrix}
\label{sec:choosing_q}

The goal of this section is twofold. First, we look at the literature on the graph Fourier transform to show how closely it 
relates to our definition, including the classical Fourier transform for temporal signals. This is summarized in 
\autoref{tab:choosing_q:gft}. The second goal is to give examples of graph Fourier transforms that use the newly introduced 
degree of freedom 
allowed by the introduction of the \qmatrix-inner product.

\begin{table*}[t]
  \caption{State of the art of the graph Fourier transform.}
  \label{tab:choosing_q:gft}
  \centering
  \def\arraystretch{1.2}
  \begin{tabu}{|rl|c|c|c|c|c|c|c|c|}
    \hhline{|==|=|=|=|=|=|=|=|=|}
    \rowfont[c]{\bfseries}
    Ref &
    Name & Directed & Weights & $\smoothness$ & \qmatrix
      & Orthon. & $\smoothness(\allone)=0$ & $\smoothness(\deltaibf{i})=cst$ & \fundmat \\
    \hline
    \cite{Shuman.IEEESP.2013} &
    Comb. Lapl. & \xmark & Non-neg. & $\lapl=\degree-\adja$ & \eye
      & \cmark & \cmark & \xmark & \lapl \\
    \hline
    \cite{Shuman.IEEESP.2013} &
    Norm. Lapl. & \xmark & Non-neg. & $\normlapl=\degree^{-\frac{1}{2}}\lapl\degree^{-\frac{1}{2}}$ & \eye
      & \cmark & \xmark & \cmark & \normlapl\\
    \hline
    \multirow{2}{*}{\cite{Sandryhaila.TSP.2013}} &
    \multirow{2}{*}{Graph Shift}
      & \xmark & \multirow{2}{*}{Complex} & $(\eye-\adja^\text{norm})^{\trconj}\times$\hspace{1.5em} & \multirow{2}{*}{$\eye$}
        & \cmark & \multirow{2}{*}{\xmark} & \multirow{2}{*}{\xmark} & \multirow{2}{*}{\adja} \\
      && \cmark & & \hspace{2em}$(\eye-\adja^\text{norm})$ & & \xmark & & & \\
    \hline
    \cite{Sardellitti.IEEEST.2017} &
    Graph Cut & \cmark & Non-neg. & $\gdv$ & \eye
      & Approx. & \cmark & \xmark & \textit{n.a.} \\
    \hline
    &
    \multirow{2}{*}{RW Lapl.}
      & \multirow{2}{*}{\xmark} & \multirow{2}{*}{Non-neg.} & $\rwlapl=\degree^{-1}\lapl$ & \eye
        & \xmark & \multirow{2}{*}{\cmark} & \multirow{2}{*}{\cmark} & \multirow{2}{*}{\rwlapl} \\
     && & & $\lapl$ & \degree & \cmark & & & \\
    \hhline{|==|=|=|=|=|=|=|=|=|}
  \end{tabu}
\end{table*}

\subsection{The Dot Product: \texorpdfstring{$\qmatrix=\eye$}{Q=I}}
\label{sec:choosing_q:dot_prod}

\subsubsection{Temporal Signals}
\label{sec:choosing_q:dot_prod:dsp}

This case derives from classical digital signal processing \cite{Oppenheim.BOOK.2013}, with $x[k]$ a periodic temporal signal of 
period $T$ and sampled with sampling frequency $N/T$ ($N$ samples per period). This sampling corresponds to a ring graph with $N$ 
vertices. In this case, energy is classically defined as a scaled $\ell_2$-norm of the vector $[x[0],\dots,x[N-1]]^T$, \ie, 
$E_\xbf=\cramped{\frac{T}{N}\xbf^{\trconj}\eye\xbf}$. DFT modes are eigenvectors of the continuous Laplacian operator 
$\Delta_T\xbf=\smash[t]{\cramped{\frac{d^2\xbf}{dt^2}}}$, thus corresponding to the variation 
$\smoothness(\xbf)=\cramped{\langle\Delta_T\xbf,\xbf\rangle=\xbf^{\trconj}\Delta_T\xbf}$. Finally, DFT modes are orthogonal \wrt 
the dot product.

\subsubsection{Combinatorial and Normalized Laplacians}
\label{sec:choosing_q:dot_prod:comb_lapl}

This classical case relies on the computations of the eigenvectors of the combinatorial Laplacian $\lapl=\degree-\adja$ or of the 
normalized Laplacian $\normlapl=\degree^{-1/2}\lapl\degree^{-1/2}$ to define the graph Fourier modes \cite{Shuman.IEEESP.2013}. 
These graph Fourier transforms are exactly the \textgftaccro{\lapl}{\eye} and the \textgftaccro{\normlapl}{\eye}.

\subsubsection{Graph Shift}
\label{sec:choosing_q:dot_prod:adja}

This is a more complex case where the (generalized) eigenvectors of the adjacency matrix are used as graph Fourier modes 
\cite{Sandryhaila.TSP.2013}. When the graph is undirected, it can be shown that this case corresponds to the 
\textgftaccro{(\eye-\adjanorm)^2}{\eye} \cite{Sandryhaila.TSP.2014}, with $\smoothness(\xbf)=\|\xbf-\adjanorm\xbf\|_2^2=
\gqv(\xbf)$, the \emph{Graph Quadratic Variation}. An 
alternative definition of smoothness based on the $\ell_1$-norm of $\xbf-\adjanorm\xbf$ 
leading to the \emph{Graph Total Variation} $\smoothness(\xbf)=\|\xbf-\adjanorm\xbf\|_1=\gtv(\xbf)$
is also investigated in 
\cite{Sandryhaila.TSP.2014}. However, this norm leads to a different graph Fourier transform when used to solve 
\autoref{eq:energy:gfm_min}, since an $\ell_1$-norm promotes sparsity (smaller support in graph Fourier modes) while an 
$\ell_2$-norm promotes smoothness. Finally, when the graph is directed, the resulting GFT of 
\cite{Sandryhaila.TSP.2013} has no guarantee that its graph Fourier modes are orthogonal. Note that the same 
$\smoothness(\xbf)=\|\xbf-\adjanorm\xbf\|_2^2$ can be used in the directed case, leading to 
$\smoothmat=(\eye-\adjanorm)^{\trconj}(\eye-\adjanorm)$. Solving \autoref{eq:energy:gfm_min} is then equivalent to computing 
the SVD of $(\eye-\adjanorm)$ and using its right singular vectors as \textgfms{\smoothmat}{\eye}. However, these differ from the 
eigenvectors of $(\eye-\adjanorm)$ used in \cite{Sandryhaila.TSP.2013} since those are orthogonal.

\subsubsection{Directed Variation Approach (\texorpdfstring{$\smoothness=\gdv$}{Delta=GDV})}
\label{sec:choosing_q:dot_prod:gdv}

We presented in \autoref{sec:energy:gdv_relation} a recently proposed GFT \cite{Sardellitti.IEEEST.2017}. 
This approach minimizes the sum of directed variation of an orthogonal set of graph 
signals (see \autoref{eq:energy:gfm_min_sum}). As explained in \autoref{sec:energy:gdv_relation}, the 
\textgftaccro{\smoothness}{\eye} is a directed GFT as described in \cite{Sardellitti.IEEEST.2017}.

\subsection{The Degree-Norm (Random Walk Laplacian): \texorpdfstring{$\qmatrix=\degree$}{Q=D}}
\label{sec:choosing_q:rwlapl}

The random walk Laplacian, $\rwlapl=\degree^{-1}\lapl$, is not widely used in the graph 
signal processing literature, given its lack of symmetry, so that its eigenvectors are not necessarily 
orthogonal. Therefore, the graph Fourier matrix \fourier is not unitary, hence naively computing it through the matrix inverse 
$\fourier=\Ubf^{-1}$ is not efficient. Yet, this normalization is successfully used in clustering in a similar manner to the 
combinatorial and normalized Laplacians \cite{Luxburg.STATCOMP.2007}. 
Noticeably, it can be leveraged to compute an optimal embedding of the vertices into a lower dimensionnal space, using 
the first few eigenvectors \cite{Belkin.NEURCOMP.2003}.
In graph signal processing we can cite 
\cite{Narang.TSP.2013,Gadde.ICIP.2013} as examples of applications of the random walk Laplacian. Another example, in image 
processing, is \cite{Liu.TIP.2017} where the authors use the random walk Laplacian as smoothness prior for soft 
decoding of JPEG images through $\xbf^{\trconj}\lapl\degree^{-1}\lapl\xbf=\|\rwlapl\xbf\|_\degree^2$.

Our framework actually allows for better insights on this case from a graph signal processing perspective%
\footnote{The property that the eigenvectors of the random walk Laplacian are orthogonal \wrt the \qmatrix-norm is well known 
in the literature. We are however the first to make the connection with a properly defined graph Fourier transform.}.
Indeed, considering the inner product with $\qmatrix=\degree$, we obtain the \textgftaccro{\lapl}{\degree}, whose fundamental 
matrix is $\fundmat=\degree^{-1}\lapl=\rwlapl$. In \cite{Liu.TIP.2017}, this leads to minimizing 
$\|\rwlapl\xbf\|_\degree^2=\|\fundmat\xbf\|_\degree^2$ which is equivalent to minimizing the energy in the higher spectrum since 
\fundmat is a high pass filter. This GFT is orthonormal \wrt the \degree-inner product, leading to a graph signal energy 
definition based on the \emph{degree-norm}: $E_\xbf=\|\xbf\|_\degree^2=\xbf^{\trconj}\degree\xbf$. As stated in 
\autoref{rem:energy:hilbert}, this normalization is related to the normalized Laplacian through the relation 
$\rwlapl=\degree^{-1/2}\normlapl\degree^{1/2}$, such that \rwlapl and \normlapl share the same eigenvalues and their eigenvectors are 
related: If $\xbf$ is an eigenvector of \rwlapl, $\hilbertmap{}(\xbf)$ is an eigenvector of \normlapl.

By combining the variation of the combinatorial Laplacian with the normalization of the normalized Laplacian, the random walk Laplacian  
achieves properties that are desirable for a sensor network: 
$\smoothness\left(\deltaibf{i}/\|\deltaibf{i}\|_\degree\right)=\frac{d_i}{d_i}=1$ since $\|\deltaibf{i}\|_\degree=d_i$, and 
$\lapl\allone=\mathbf{0}=0\cdot\degree\allone$ \ie, the graph frequencies are normalized, the constant signals have 
zero variation, and all impulses have the same variation, while having different energies%
\footnote{Impulse energies show here the irregularity of the graph structure, and as such are naturally not constant.}.

This case is actually justified in the context of manifold sampling in the extension \cite{Hein.JMLR.2007} of 
\cite{Belkin.JCSS.2008}. The authors show that under some conditions on the weights of the edges of the similarity graph, the 
random walk Laplacian is essentially the continuous Laplacian, without additive or multiplicative term, even if the probability 
density function of the samples is not uniform.

\subsection{Bilateral Filters: \texorpdfstring{$\qmatrix=\eye+\degree$}{Q=I+D}}
\label{sec:choosing_q:bf}

Bilateral filters are used in image processing to denoise images while retaining clarity on edges in the image 
\cite{Tomasi.ICCV.1998}:
\[
  y_i = \frac{1}{1+d_i} \left(x_i + \sum_jw(ij)x_j\right)
\]
with
\[
  w(ij) =
      \exp\left(-\frac{\|p_i - p_j\|^2}{2\sigma_\text{d}^2}\right)
      \exp\left(-\frac{\|I(i) - I(j)\|^2}{2\sigma_\text{i}^2}\right)
\]
where $p_i$ is the position of pixel $i$, $I(i)$ is its intensity, and $\sigma_\text{d}$ and $\sigma_\text{i}$ are parameters. 
Intuitively, weights are smaller when pixels are either far (first Gaussian kernel) or of different intensities (second Gaussian 
kernel). This second case corresponds to image edges. We can rewrite this filtering operation in matrix form as:
\[
  \ybf = \left(\eye+\degree\right)^{-1}(\eye + \adja)\xbf
    = \eye - \left(\eye+\degree\right)^{-1}\lapl\xbf
  \text{.}
\]
In other words, using $\fundmat=\left(\eye+\degree\right)^{-1}\lapl$, we obtain that bilateral filtering is the polynomial graph 
filter $\eye-\fundmat$. This corresponds exactly to a graph filter with the \textgftaccro{\lapl}{\eye+\degree}. 

Moreover, given a noisy observation $\ybf=\xbf+\nbf$ of the noiseless image $\xbf$, noise on the output of this filter is given 
by $(\eye-\fundmat)\ybf-\xbf=\fundmat\xbf+(\eye-\fundmat)\nbf$. This noise can be studied using the $(\eye+\degree)$-MSE 
introduced in \autoref{sec:filters:mse}. Indeed, we can experimentally observe that pixels do not contribute equally to the 
overall error, with pixels on the edges (lower degree) being less filtered than pixels in smooth regions (higher degree). 
experimentally, we observe that $(\eye-\fundmat)\nbf$ is an $(\eye+\degree)$-WN, thus validating the use of the 
$(\eye+\degree)$-MSE. This approach of quantifying noise is coherent with human perception of noise as the human eye is more 
sensitive to small changes in smooth areas. We will develop the study of bilateral filters with respect to our setting in a 
future communication.

Finally, note that the expression of bilateral filtering shown here is different than in \cite{Gadde.ICIP.2013}:
\[
  y_i = \frac{1}{d_i} \sum_jw(ij)x_j
  \text{.}
\]
This expression is equivalent to considering the filter $\eye-\fundmat$ with the \textgftaccro{\lapl}{\degree}, and is not the 
original approach of \cite{Tomasi.ICCV.1998}.

\subsection{Voronoi Cell Areas: \texorpdfstring{$\qmatrix=\Cbf$}{Q=C}}
\label{sec:choosing_q:voronoi}

In the motivating example of a sensor network, we wished to find a graph Fourier transform that is not biased by the particular 
sampling being performed. Considering the energy of the signal being measured, this energy should not vary with the sampling. In 
the continuous domain, we observe that energy of a continuous signal $\widetilde{s}$ is defined as:
\[
  E_s:=\int|\widetilde{s}(x)|^2dx
  \text{.}
\]
In the definition above, $dx$ acts as an elementary volume, and the integral can be interpreted as a sum over elementary volumes 
of the typical value $|\widetilde{s}(x)|$ within that volume times the size of the volume $dx$. The discrete version of this 
integral is therefore:
\[
  E_s\approx \sum_i \left|\widetilde{s}(x_i)\right|^2\vol(i)=\sum_i \left|s_i\right|^2\vol(i)
  \text{,}
\]
with \sbf the sampled signal on the points $\{x_i\}_i$.

Given a particular sampling, the question is then what is a good value of $\vol(i)$? A simple and intuitive approximation is to 
approximate this volume with the volume of the subspace of points whose closest sample is $x_i$. This subspace is exactly the 
Voronoi cell of $i$, and the volume is therefore the area of the cell in 2D. Let $c_i$ be this area, and 
$\Cbf=\diag(c_1,\dots,c_N)$. We obtain:
\[
  E_s\approx \sum_i \left|s_i\right|^2c_i=\sbf^{\trconj}\Cbf\sbf=\|\sbf\|_\Cbf^2
  \text{.}
\]
Intuitively, this corresponds to interpolating the sampled signal into a piecewise constant signal for which the signal values 
are equal within each Voronoi cell, and then computing the continuous energy of this interpolated signal. Other interpolation 
schemes could be considered. 
For example, if we assume a weighted linear interpolation scheme, then we obtain the interpolated signal 
$\widetilde{\sbf}(\xbf)=\sum_i f(\xbf,\xbf_i)s_i$ and its energy is:
\[
  E_{\widetilde{\sbf}}
    =\int\left|\widetilde{\sbf}(\xbf)\right|^2d\xbf
    =\sum_{i,j}\left(\int f(\xbf,\xibf{i})^{\trconj}f(\xbf,\xibf{j})d\xbf\right)s_i^{\trconj}s_j
  \text{,}
\]
and we have $E_{\widetilde{\sbf}}=\left\|\sbf\right\|_\qmatrix^2$ 
with $q_{ij}=\int f(\xbf,\xibf{i})^{\trconj}f(\xbf,\xibf{j})d\xbf$. As soon as there is one location \xbf whose interpolation
involves two samples $i$,$j$, then this interpolation scheme corresponds a non-diagonal matrix \qmatrix.
However, as we will show in 
\autoref{sec:experiments:sensor_nets}, the approach based on the Voronoi cell areas gives already good results 
compared to the state of the art of graph signal processing for sensor networks.

\section{Experiments}
\label{sec:experiments}

The choice of a graph Fourier transform, by selecting a graph variation operator $\smoothness$ and a matrix \qmatrix such as 
those discussed in \autoref{sec:choosing_q}, is application dependent. In particular, meaningful properties for the graph Fourier 
transform can be drastically different. For example, in a clustering context, the practitioner is interested in extracting the 
structure, \ie, in identifying tight groups of vertices. This is in constrast to sensor networks where the particular arrangement 
of sensors should have as little influence over the results as possible, \ie, tight groups of stations should not be a source of 
bias (\eg, weather stations).

For this reason, there is no unique way of studying how good a particular graph Fourier transform is. In this section, we use 
these two applications, \ie, clustering and sensor networks, to show how our framework can be leveraged to achieve their 
application-specific goals.

All experiments were performed using the \texttt{GraSP} toolbox for \texttt{Matlab} \cite{Girault.ICASSPDEMO.2017}.

\subsection{Clustering}
\label{sec:experiments:clustering}

\input{clustering_graph.tex}

\input{clustering_clusters.tex}

In this section, we study the problem of clustering defined as 
grouping similar objects into groups of object that are similar while being dissimilar to objects 
in other groups. In a graph setting, we are interested in grouping vertices of a graph such that there are many edges within each 
group, while groups are connected by few edges.
Our goal is not to provide a new approach to perform clustering, but rather use this problem as a showcase for how using 
a well defined matrix \qmatrix can help achieve the goals of a target application.

In the context of clustering, spectral clustering extracts groups using the spectral 
properties of the graph~\cite{Luxburg.STATCOMP.2007}. More precisely, using $C$-means ($k$-means with $k=C$) on the first $C$ 
graph Fourier modes yields interesting partitions. In \cite{Luxburg.STATCOMP.2007}, the author interprets this approach using 
graph cuts, for \textgftaccro{\lapl}{\eye} (combinatorial Laplacian), \textgftaccro{\normlapl}{\eye} (normalized Laplacian) and 
\textgftaccro{\lapl}{\degree} (random walk Laplacian). We extend this interpretation here to any variation operator $\smoothness$ 
and any diagonal innner product matrix \qmatrix.

For each cluster $c$, let $\mathcal{V}_c\subset\mathcal{V}$ be the subset of its vertices. Then the set $\{\mathcal{V}_c\}_c$ of 
all these subsets is a partition of the set of vertices $\mathcal{V}$. Let the normalized cut of $\{\mathcal{V}_c\}_c$ associated 
to the \textgftaccro{\smoothness}{\qmatrix} be:
\[
  \ncutparam{\smoothness}{\qmatrix}\left(\{\mathcal{V}_c\}_c\right) := \sum_c\smoothness\left(\hbf^{(\qmatrix)}_c\right)
\]
with $\hbf^{(\qmatrix)}_c$ the normalized indicator function of cluster $\mathcal{V}_c$ verifying:
\[
  \hbf^{(\qmatrix)}_c:=\frac{1}{\|\allone_{\mathcal{V}_c}\|_\qmatrix}\allone_{\mathcal{V}_c}
\]
and $\allone_{\mathcal{V}_c}$ the indicator function of cluster $c$. This extends the normalized cut interpretation of 
\cite{Luxburg.STATCOMP.2007}. Using the notation $\Hbf=[\hbf^{(\qmatrix)}_1 \cdots \hbf^{(\qmatrix)}_C]^T$, and since \qmatrix is 
diagonal, we obtain the orthonormality property $\Hbf^{\trconj}\qmatrix\Hbf=\eye$. Finding a partition of vertices minimizing the 
normalized cut $\ncutparam{\smoothness}{\qmatrix}\left(\{\mathcal{V}_c\}_c\right)$ is therefore equivalent to finding orthonormal 
normalized indicator functions of minimal variation.

Spectral clustering is performed by first relaxing the constraint that the graph signals $\hbf^{(\qmatrix)}_c$ are normalized 
indicator functions. Using \autoref{prop:energy:gfm_min_sum_solution}, the first $C$ \textgfms{\smoothness}{\qmatrix} are 
solutions to this relaxed optimization problem. The final partition is then obtained through $C$-means on the spectral features, 
where the feature vector of vertex $i$ is $\bigl[[\ubf_0]_i,\dots,[\ubf_{C-1}]_i\bigr]^T$.

\begin{table}[b]
  \caption{$F_1$ score of the sparse (left) cluster on \autoref{fig:experiments:clustering:graph} for the clustering results of 
    \autoref{fig:experiments:clustering}.}
  \label{tab:experiments:clustering:f1}
  \centering
  \def\arraystretch{1.2}
  \begin{tabu}{|l|c|c|}
    \hhline{|=|=|=|}
    \rowfont[c]{\bfseries}
    Approach & Accuracy & Sparse $F_1$ score \\
    \hline
    $(\lapl,\eye)$ & $91.21\%$ & $6.45\%$ \\
    $(\normlapl,\eye)$ & $93.64\%$ & $46.15\%$ \\
    $(\normlapl,\eye)$, feat. norm. & $46.67\%$ & $4.35\%$ \\
    $(\lapl,\degree)$ & $\mathbf{96.67}\%$ & $\mathbf{77.55}\%$ \\
    $(\lapl,\Cbf)$ & $91.21\%$ & $6.45\%$ \\
    $(\|\xbf-\adjanorm\xbf\|_1,\eye)$ & $86.06\%$ & $4.17\%$ \\
    \hhline{|=|=|=|}
  \end{tabu}
\end{table}

The accuracy of spectral clustering is therefore driven by the choice of variation operator $\smoothness$ and inner product 
matrix $\qmatrix$. To illustrate this, and the statement that choosing \qmatrix and $\smoothness$ is application-dependent, we 
look at several choices in the context of a 2-class clustering problem with skewed clusters. Combined with an irregular 
distribution of inputs within each cluster, we are in the context of irregularity where our framework thrives through its 
flexibility.

The resulting graph is shown on \autoref{fig:experiments:clustering:graph}, with a sparse cluster on the left (30 inputs), and a 
dense one on the right (300 inputs). 
Each cluster is defined by a 2D Gaussian distribution, with an overlapping support, and samples are drawn from these
two distributions. 
We already observe that inputs are irregularly distributed, and some inputs are almost 
isolated. These almost isolated inputs are important as they represent outliers. We will see that correctly choosing \qmatrix 
alleviates the influence of those outliers. The graph of inputs is built in a $K$-nearest neighbor fashion ($K=10$) with weights 
chosen using a Gaussian kernel of the Euclidean distance ($\sigma=0.4$).

\Autoref{fig:experiments:clustering} shows the results of spectral clustering on this graph, using several GFTs
(first two rows), the analysis of $\hbf^{\smash{(\qmatrix)}}_{\text{blue}}$ in the spectral domain (third row),
and an example of \qmatrix-MSE of several ideal low pass filters on the same indicator function. 
First of all, 
the \emph{Random Walk Laplacian} case based on the \textgftaccro{\lapl}{\degree} gives the best results, consistent with 
\cite{Luxburg.STATCOMP.2007} where the author advocates 
this approach. \cite[Prop. 5]{Luxburg.STATCOMP.2007} gives an intuition on why this case works best using a random walk 
interpretation: If a starting vertex is chosen at random from the stationary distribution of the random walk on the graph, then 
the \ncutparam{\lapl}{\degree} is exactly the probability to jump from one cluster to another. Minimizing this cut, is then 
minimizing transitions between clusters.
In the spectral domain, we see that the indicator function has a lot of energy in the first few spectral components, 
with a rapid decay afterwards. We notice also, a slight increase of energy around graph frequency 1 due to the red vertices
of the sparse (red) cluster which are in the middle of the dense (blue) cluster (see 
\autoref{fig:experiments:clustering:graph:classes}). As expected, the \qmatrix-MSE decreases when the cut-off frequency 
increases. But interestingly, if we use the classical \eye-MSE, it is not decreasing with the cut-off frequency because
of the use of the \qmatrix-norm instead of the $\ell_2$-norm (see \autoref{fig:experiments:clustering_MSE_I:rwlapl}.

The combinatorial Laplacian case, \ie, using the \textgftaccro{\lapl}{\eye}, suffers from the presence of outliers. Indeed, if 
$i$ is such an outlier, then having it in a separate cluster, \ie, $\mathcal{V}_1=\{i\}$ and 
$\mathcal{V}_2=\mathcal{V}\backslash\{i\}$ for $C=2$, leads to $\smoothness(\hbf^{(\eye)}_1)=\smoothness(\hbf^{(\eye)}_2)=d_i$. 
Since $i$ is isolated, it has extremely low degree, and the resulting cut is small, making it a good clustering according to the 
\ncutparam{\lapl}{\eye}. Not accounting for irregularity of the degree in the definition of the normalized indicator function 
$\hbf^{(\eye)}_c$ leads therefore to poor clustering in the presence of outliers. A similar behavior is observed for the 
\textgftaccro{\lapl}{\Cbf} with a normalization by the Voronoi cell area that is large for isolated vertices, hence an even 
smaller normalized cut.
In the spectral domain, more Fourier components of the normalized indicator function are large, especially around the
graph frequency 0. This results in a \qmatrix-MSE that is large for our ideal low-pass filters. In other words, more lowpass 
components are required to correctly approximate the indicator function using a low-pass signal.

Next, is the normalized Laplacian case, \ie, using the \textgftaccro{\normlapl}{\eye}. 
Just as the Random Walk Laplacian case, singleton clusters with outliers are not associated to small cuts since 
$\ncutparam{\normlapl}{\eye}(\{i\},\mathcal{V}\backslash\{i\})=\ncutparam{\lapl}{\degree}(\{i\},\mathcal{V}\backslash\{i\})$. 
However, a careful study of the graph Fourier modes reveals the weakness of this case. Indeed, using the well known relation 
between the \textgfms{\normlapl}{\eye} $\{\ulbf{l}\}_l$ and the \textgfms{\lapl}{\degree}$\{\vlbf{l}\}_l$ with 
$\hilbertmap{}(\vlbf{l})=\ulbf{l}=\degree^{1/2}\vlbf{l}$, this case deviates from the Random Walk Laplacian by introducing a 
scaling of graph Fourier modes by the degree matrix. 
Differences are then noticeable in the 
spectral feature space (bottom plots in \autoref{fig:experiments:clustering}). Indeed, the graph Fourier modes verify 
$[\ulbf{l}]_i=\sqrt{d_i}[\vlbf{l}]_i$, such that low degree vertices, \eg, outliers and vertices in the boundary of a cluster, 
have 
spectral features of smaller magnitude: $\bigl\|\bigl[[\ulbf{0}]_i, [\ulbf{1}]_i\bigr]^T\bigr\|\ll \bigl\|\bigl[[\vlbf{0}]_i, 
[\vlbf{1}]_i\bigr]^T\bigr\|$. In other words, spectral features 
of low degree vertices are moved closer to the origin of the 
feature space, and closer to each other. $C$-means cluster those vertices together, resulting in the poor 
separation of clusters we see on \autoref{fig:experiments:clustering:normlapl}.
In the spectral domain, we clearly see the interest of using the random walk Laplacian approach: the normalized 
indicator functions are not lowpass. This is further seen on the \qmatrix-MSE that decreases continuously with the cutoff frequency
without the fast decay we wish for.

\input{clustering_MSE_I.tex}

The spectral clustering literature is actually not using the raw spectral features to run $C$-means, but normalize those features 
by their $\ell_2$-norm prior to running $C$-means \cite{Ng.ANIPS.2002}. This is actually equivalent to the projection of the 
\textgfms{\lapl}{\degree} $\{\vlbf{l}\}_l$ on the unit circle:
\begin{align*}
  [\widetilde{\ulbf{0}}]_i &= \cos\left(\theta_i\right) &
  [\widetilde{\ulbf{1}}]_i &= \sin\left(\theta_i\right) &
  \theta_i &= \atan\Bigl(\|\allone\|_\degree[\vlbf{1}]_i\Bigr)
  \text{.}
\end{align*}
We can use this relation to characterize how spectral features of vertices are modified compared to the 
\textgftaccro{\lapl}{\degree} approach, and how $C$-means behaves. Indeed, separation of features (\ie, how far the spectral 
features of two vertices are) is modified by the transformation above. Looking at the norm of the gradient of these spectral 
features 
yields:
\[
  \left\|\overrightarrow{\grad}
  \begin{bmatrix}[\widetilde{\ulbf{0}}]_i \\ [\widetilde{\ulbf{1}}]_i\end{bmatrix}
  \right\|_2
  = \frac{\|\allone\|_\degree}{1+\left(\|\allone\|_\degree[\vlbf{1}]_i\right)^2}
\]
Therefore, the gradient is smaller for higher values of $|[\vlbf{1}]_i|$: This spectral feature normalization brings features 
closer for higher values of $|[\vlbf{1}]_i|$. As features of the dense and sparse clusters are identified by the magnitude of 
$|[\vlbf{1}]_i|$, this normalization actually brings features of the two clusters closer. This yields less separation in the 
feature space and poor clustering output from $C$-means (see \autoref{fig:experiments:clustering:normlapl_orig}).

Finally, we performed the same specral clustering approach using the \textgftaccro{\smoothness}{\eye} with 
$\smoothness(x)=\|\xbf-\adjanorm\xbf\|_1$ the graph total variation. Results on \autoref{fig:experiments:clustering:a1} show that 
this approach actually identifies the stronger small clusters rather than larger clusters. More importantly, in this case, the 
spectral features of both clusters are not separated, such that $C$-means cannot separate clusters based on the spectral 
features. Furthermore, it turns out that the clusters given by these spectral features are not the best with an 
$\ncutparam{\gtv}{\eye}$ of $8.21$, while those given by the combinatorial Laplacian yields a $\ncutparam{\gtv}{\eye}$ of $6.61$. 
In other words, the relaxation we made that consider arbitrary $\hbf^{(\eye)}_c$ instead of normalized indicator functions 
leads to large errors.
The spectral domain analysis suffers from the difficulty to compute efficiently the graph Fourier modes, since the 
graph variation is not HPSD. However, the first few components of the normalized indicator function show that this signal
is now low pass, with a lot of energy in these components. \qmatrix-MSE is not sharply decreasing either and stays very high.

\subsection{Sensor Networks}
\label{sec:experiments:sensor_nets}

Here we explore our motivating example of a sensor network, where the goal is the opposite of that in the one 
above: clusters of vertices should not bias the results. More precisely, we are interested in studying signals measured by sensor 
networks where the location of the sensors is unrelated to the quantity measured, \eg, temperature readings in a weather station 
network. In this section, we show that we can obtain a good definition of inner product achieving this goal.

\input{sensor_net_figures.tex}

Let $\mathcal{G}^{(\text{U})}$ be a graph whose vertices are uniformly sampled on a 2D square plane, and 
$\mathcal{G}^{(\text{NU})}$ a similar graph obtained with a non-uniform sampling distribution (see 
\autoref{fig:experiments:sensor_net:energy:non_unif:degree} where the areas of higher density are indicated). Edges of both 
graphs are again selected using $K$-nearest neighbors, with $K=10$. Weights are chosen as a Gaussian kernel of the Euclidean 
distance $w_{ij}=\exp\left(-\dist(i,j)^2/(2\sigma^2)\right)$, with $\sigma=0.3$ (empirically chosen to have a good spread of 
weights). Example graphs with 500 sampled vertices are shown in 
\autorefs{fig:experiments:sensor_net:energy:unif:degree,fig:experiments:sensor_net:energy:non_unif:degree}. In this experiment, 
we are interested in studying how we can remove the influence of the particular arrangement of sensors such as to obtain the same 
results from one graph realization to another. To that end, we generate 500 graph realizations, each with 1000 vertices, to 
compute empirical statistics.

The signals we consider are pure sine waves, which allows us to control the  
energies and frequencies of the signals we use. We experiment with several signals for each frequency and report the worst case 
statistics over those signals of equal (continuous) frequency. This is done by choosing several values of phase shift 
for each frequency.

Let $\widetilde{s_\nu}(x,y;\varphi)=\cos(2\pi\nu x_i+\varphi)$ be a continuous horizontal Fourier mode of frequency $\nu$ phase 
shifted by $\varphi$. For a given graph $\mathcal{G}$ generated with any of the two schemes above, we define 
$s_\nu(i;\varphi,\mathcal{G})=\widetilde{s_\nu}(x_i,y_i;\varphi)$ the sampled signal on the graph. Its energy \wrt the 
\qmatrix-norm is then:
\[
  E_\nu^{(\qmatrix)}(\varphi;\mathcal{G}):=\left\|s_\nu(\_;\varphi,\mathcal{G})\right\|_\qmatrix^2
  \text{.}
\]
Note that this energy depends on the graph if \qmatrix does, \ie, we have $\qmatrix(\mathcal{G})$ which may vary from one graph 
to another such as $\qmatrix(\mathcal{G})=\degree(\mathcal{G})$. We use the shorter notation \qmatrix to keep notations shorter.

To get better insights on the influence of the sensor sampling, we are interested in statistical quantities of the graph signal 
energy. First, we study the \emph{empirical mean}:
\[
  \mu^{(\qmatrix)}(\nu;\varphi)
    := \left\langle E_\nu^{(\qmatrix)}(\varphi;\mathcal{G})\right\rangle_\mathcal{G}
    = \frac{1}{N_\mathcal{G}}\sum_{n=1}^{N_\mathcal{G}} E_\nu^{(\qmatrix)}(\varphi;\mathcal{G}_n)
  \text{,}
\]
with $N_\mathcal{G}$ the number of sampling realizations. It is interesting to study how the mean varies 
depending on $\varphi$ and $\nu$ in order to 
check whether the graph signal energy remains constant, given that the energy of the continuous signals being sampled is constant.  
The first thing we observe is that the averaging over $\varphi$, \ie, over many signals of equal continuous frequency, yields 
the same average mean for all continuous frequencies. However, $\mu^{(\qmatrix)}(\nu;\varphi)$ does depend on $\varphi$. To show 
this, we use the \emph{maximum absolute deviation} of the normalized mean 
$\bar{\mu}^{(\qmatrix)}(\nu;\varphi):=\mu^{(\qmatrix)}(\nu;\varphi)/\langle \mu^{(\qmatrix)}(\nu;\varphi)\rangle_\varphi$:
\[
  m^{(\qmatrix)}(\nu)
    :=\max_\varphi\left|1-\bar{\mu}^{(\qmatrix)}(\nu;\varphi)\right|
  \text{.}
\]
Use of the normalized mean is necessary to be able to compare the various choices of \qmatrix since they can yield quite different 
average mean $\langle \mu^{(\qmatrix)}(\nu;\varphi)\rangle_\varphi$.

However, the mean is only characterizing the bias of the graph signal energy approximation to the continuous signal energy. We 
also need to characterize the variance of this estimator. To that end, we consider the \emph{empirical standard deviation} of the 
graph signal energy:
\[
  \sigma^{(\qmatrix)}(\nu;\varphi)
    := \sqrt{\frac{1}{N_\mathcal{G}-1}\sum_{n=1}^{N_\mathcal{G}} 
    \left(E_\nu^{(\qmatrix)}(\varphi;\mathcal{G}_n)-\mu^{(\qmatrix)}(\nu;\varphi)\right)^2}
  \text{.}
\]
This quantity shows how much the actual sampling being performed influences the signal energy estimator. Since the mean energy is 
influenced by the choice of \qmatrix, we need to normalize the standard deviation to be able to compare results between various 
choices of \qmatrix. This yields the \emph{coefficient of variation}, and we report its maximum over all signals of equal 
(continuous) frequency:
\[
  \CV^{(\qmatrix)}(\nu):=\max_\varphi \frac{\sigma^{(\qmatrix)}(\nu;\varphi)}{\mu^{(\qmatrix)}(\nu;\varphi)}
  \text{,}
\]
from which we can study the variance of the graph signal energy estimator depending on the continuous frequency.

We experiment here with three choices for \qmatrix. The first one a standard GFT from the literature,  
corresponding to the $\ell_2$-norm: $\qmatrix=\eye$. The second one corresponds to the random-walk Laplacian  with 
$\qmatrix=\degree$. The third one is our novel approach based on the Voronoi cell area inner product $\qmatrix=\Cbf$ (see 
\autoref{sec:choosing_q:voronoi}).

Results for $\mathcal{G}^{(\text{U})}$ and $\mathcal{G}^{(\text{NU})}$ are given in \autoref{fig:experiments:sensor_net:energy}. 
The coefficient of variation shows here the strong advantage of using the Voronoi cell area inner product: 
$\cramped{\CV^{(\Cbf)}(\nu)}$ is very small in the lower continuous spectrum, and almost zero (see 
\autorefs{fig:experiments:sensor_net:energy:unif:energy_cv,fig:experiments:sensor_net:energy:non_unif:energy_cv}). In other 
words, for signals that do not vary too quickly compared to the sampling density, \emph{the $\Cbf$-norm gives an estimated  
energy with small variance, removing the influence of the particular arrangement of sensors}. This is true for both the uniform 
distribution and the non-uniform distribution. Both the dot product $\qmatrix=\eye$ and the degree norm $\qmatrix=\degree$ have 
larger variance here.

Finally, the maximum absolute deviation on 
\autorefs{fig:experiments:sensor_net:energy:unif:energy_deviation,fig:experiments:sensor_net:energy:non_unif:energy_deviation} 
shows again the advantage of the \Cbf-norm. Indeed, considering several signals of equal continuous frequencies, these should 
have equal average energy. While this is the case for both the dot product and the \Cbf-norm in the lower spectrum for a uniform 
sampling, using a non-uniform sampling yields a strong deviation for the dot product. The \Cbf-norm appears again as a good 
approximation to the continuous energy with small deviation between signals of equal frequencies, and between samplings.

\begin{remark}
  In \cite{Coifman.ACHA.2006.Maps,Hein.JMLR.2007}, the authors advocate for the use of the random walk Laplacian as a good 
  approximation for the continuous (Laplace-Beltrami) operator of a manifold. We showed here that this case is not ideal when it 
  comes to variance of the signal energy estimator, but in those communications, the authors actually normalize the weights of 
  the graph using the degree matrix prior to computing the random walk Laplacian, thus working on a different graph. If 
  $\widetilde{\adja}$ is the adjacency prior its normalization, and $\widetilde{\degree}$ the associated degree matrix, then 
  $\adja=\widetilde{\degree}^{-1}\widetilde{\adja}\widetilde{\degree}^{-1}$. This normalization is important when we look at the 
  degree. Indeed, without it, we see on 
  \autoref{fig:experiments:sensor_net:energy:unif:degree}-\subref{fig:experiments:sensor_net:energy:unif:voronoi} and 
  \ref{fig:experiments:sensor_net:energy:non_unif:degree}-\subref{fig:experiments:sensor_net:energy:non_unif:voronoi} that degree 
  and Voronoi cell area evolve in opposed directions: large degrees correspond to small areas. Pre-normalizing by the degree 
  corrects this. Using this normalization leads to better results with respect to the energy (omitted here to save space), 
  however, Voronoi cells still yield the best results.
\end{remark}

These results show that it is possible to better account for irregularity in the definition of the energy of a graph signal: 
The Voronoi cell area norm yields very good results when it comes to 
analysing a signal lying in a Euclidean space (or on a manifold) independently of the sampling performed.

\section{Conclusion}

We showed that it is possible to define an orthonormal graph Fourier transform with respect to any inner product. This allows in 
particular to finely account for an irregular graph structure. Defining the associated graph filters is then straightforward 
using the fundamental matrix of the graph. Under some conditions, this led to a fundamental matrix that is easy to compute and 
efficient to use. We also showed that the literature on graph signal processing can be interpreted with this graph Fourier 
transform, often with the dot product as inner product on graph signals. Finally, we showed that we are able to obtain promising 
results for sensor networks using the Voronoi cell areas inner 
product.

This work calls for many extensions, and giving a complete list of them would be too long for this conclusion. At this time, 
we are working on the sensor network case to obtain graph signal energies with even less variance, especially once the signals 
are filtered. We are also working on studying bilateral filters, and image smoothers in general \cite{Milanfar.SPMAG.2013}, 
and extending them to graph signals with this new setting. We explore also an extension of the study of random graph signals and 
stationarity with respect to this new graph Fourier transform. Finally, many communications on graph signal processing can be 
re-interpreted using alternative inner product that will give more intuitions on the impact of irregularity.

\section*{Acknowledgement}

The authors would like to thank the anonymous reviewers for their careful reading and constructive comments that helped
improve the quality of this communication.

\appendix

\subsection{Low Pass Filtering of an Indicator Function}
\label{sec:lowpass_filter_clustering}

\input{clustering_lowpass}

We now show an example to compare the quality of different low pass approximations to the normalized cluster indicator function, 
under different GFT definitions. This is motivated by the fact that spectral clustering techniques make use of the low frequencies 
of the graph spectrum. \autoref{fig:experiments:clustering_indicator_low_pass} shows these approximations in the vertex domain. 
Note that these correspond to the \qmatrix-MSE plots of \autoref{fig:experiments:clustering}.

Comparing results for $l=1$ and $l=5$ we note that for both \lapl and \normlapl, the low pass approximation is good, and 
significantly better than that achieved based on the GFT corresponding to graph total variation. However, the 
\textgftaccro{\lapl}{\eye} suffers from isolated vertices ($l=1$ case on \autoref{fig:experiments:clustering_indicator_low_pass}), 
and the approximation needs more spectral components for an output closer to the indicator function ($l=5$ case), while the 
\textgftaccro{\normlapl}{\eye} is clearly biased with the degree (smaller amplitudes on the vertices of the cluster boundaries).

This confirms that the indicator function approximation (and thus spectral clustering performance) is better for the 
\textgftaccro{\lapl}{\degree}, which does not have the bias towards isolated vertices shown by the \textgftaccro{\lapl}{\eye} 
(impulses on those isolated vertices are associated to lowpass signal, while they are associated to the graph frequency 1 for the 
\textgftaccro{\lapl}{\degree}), and without the bias of the first Fourier mode towards vertex degrees of the 
\textgftaccro{\normlapl}{\eye}.

\bibliographystyle{IEEEbib}
\bibliography{bibliography}

\end{document}

%% file: ring_graph.tex
\begin{figure*}[t]
  \centering
  \setlength\tabcolsep{1.3pt}
    \begin{tabular}{rcccccccccccc}
    &
    \tiny $\lambda_0=\input{data/ring_Q0_1_mode_1_eigval.tex}$
    &
    &
    \tiny $\lambda_1=\input{data/ring_Q0_1_mode_2_eigval.tex}$
    &
    \tiny $\lambda_2=\input{data/ring_Q0_1_mode_3_eigval.tex}$
    &
    \tiny $\lambda_3=\input{data/ring_Q0_1_mode_4_eigval.tex}$
    &
    &
    \tiny $\lambda_4=\input{data/ring_Q0_1_mode_5_eigval.tex}$
    &
    \tiny $\lambda_5=\input{data/ring_Q0_1_mode_6_eigval.tex}$
    &
    &
    \tiny $\lambda_6=\input{data/ring_Q0_1_mode_7_eigval.tex}$
    &
    \tiny $\lambda_7=\input{data/ring_Q0_1_mode_8_eigval.tex}$
    \\
    \rotatebox{90}{\footnotesize \hspace{0.05cm}$\smash{q_1=0.1}$}
    &
    \includegraphics{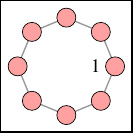}
    &
    &
    \includegraphics{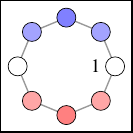}
    &
    \includegraphics{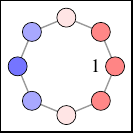}
    &
    \includegraphics{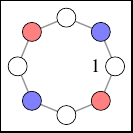}
    &
    &
    \includegraphics{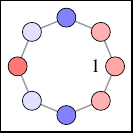}
    &
    \includegraphics{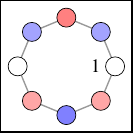}
    &
    &
    \includegraphics{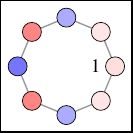}
    &
    \includegraphics{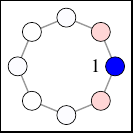}
    &
    \includegraphics{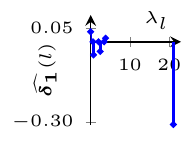}
    \\[0.3cm]
    &
    \tiny $\lambda_0=\input{data/ring_I_mode_1_eigval.tex}$
    &
    &
    \tiny $\lambda_1=\input{data/ring_I_mode_2_eigval.tex}$
    &
    \tiny $\lambda_2=\input{data/ring_I_mode_3_eigval.tex}$
    &
    \tiny $\lambda_3=\input{data/ring_I_mode_4_eigval.tex}$
    &
    \tiny $\lambda_4=\input{data/ring_I_mode_5_eigval.tex}$
    &
    &
    \tiny $\lambda_5=\input{data/ring_I_mode_6_eigval.tex}$
    &
    \tiny $\lambda_6=\input{data/ring_I_mode_7_eigval.tex}$
    &
    \tiny $\lambda_7=\input{data/ring_I_mode_8_eigval.tex}$
    \\
    \rotatebox{90}{\footnotesize \hspace{0.1cm}$\smash{\qmatrix=\eye}$}
    &
    \includegraphics{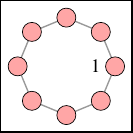}
    &
    &
    \includegraphics{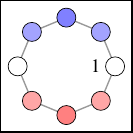}
    &
    \includegraphics{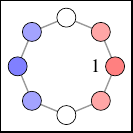}
    &
    \includegraphics{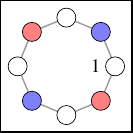}
    &
    \includegraphics{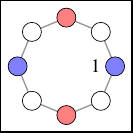}
    &
    &
    \includegraphics{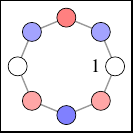}
    &
    \includegraphics{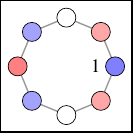}
    &
    \includegraphics{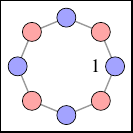}
    &
    &
    \includegraphics{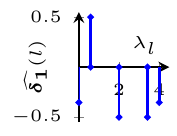}
    \\[0.3cm]
    &
    \tiny $\lambda_0=\input{data/ring_Q10_mode_1_eigval.tex}$
    &
    \tiny $\lambda_1=\input{data/ring_Q10_mode_2_eigval.tex}$
    &
    \tiny $\lambda_2=\input{data/ring_Q10_mode_3_eigval.tex}$
    &
    \tiny $\lambda_3=\input{data/ring_Q10_mode_4_eigval.tex}$
    &
    \tiny $\lambda_4=\input{data/ring_Q10_mode_5_eigval.tex}$
    &
    &
    \tiny $\lambda_5=\input{data/ring_Q10_mode_6_eigval.tex}$
    &
    \tiny $\lambda_6=\input{data/ring_Q10_mode_7_eigval.tex}$
    &
    &
    \tiny $\lambda_7=\input{data/ring_Q10_mode_8_eigval.tex}$
    \\
    \rotatebox{90}{\footnotesize \hspace{0.05cm}$\smash{q_1=10}$}
    &
    \includegraphics{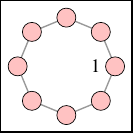}
    &
    \includegraphics{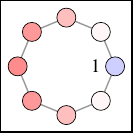}
    &
    \includegraphics{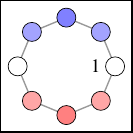}
    &
    \includegraphics{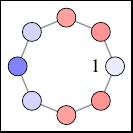}
    &
    \includegraphics{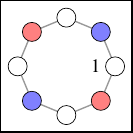}
    &
    &
    \includegraphics{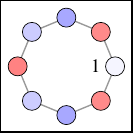}
    &
    \includegraphics{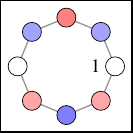}
    &
    &
    \includegraphics{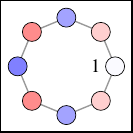}
    &
    &
    \includegraphics{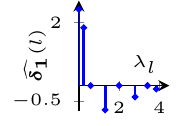}
  \end{tabular}
  \caption{\textgfms{\lapl}{\qmatrix} of the ring graph with $N=8$ vertices. $\qmatrix=\diag(q_1,1,\dots,1)$ with 
    one choice of $q_1$ per row of the table, for increasing values of $q_1$. Modes are not aligned on their index, but when 
    they correspond either exactly
    (3$^\text{rd}$, 5$^\text{th}$, and 8$^\text{th}$ column), or approximately, to highlight the impact 
    of $q_1$ on the graph Fourier modes. Colors scale from deep blue for $-1$ to deep red for $1$ and color being lighter as
    values get closer to 0 until white for 0. Note that $[\ulbf{7}]_1=-3.08$ for $q_1=0.1$ is out of this range.
    The last column shows the graph Fourier transform $\widehat{\deltaibf{1}}$ of the impulse \deltaibf{1}.
  }
  \label{fig:ring_graph}
\end{figure*}

%% file: data/ring_Q0_1_mode_1_eigval.tex
0

%% file: data/ring_Q0_1_mode_2_eigval.tex
0.59

%% file: data/ring_Q0_1_mode_3_eigval.tex
0.74

%% file: data/ring_Q0_1_mode_4_eigval.tex
2

%% file: data/ring_Q0_1_mode_5_eigval.tex
2.42

%% file: data/ring_Q0_1_mode_6_eigval.tex
3.41

%% file: data/ring_Q0_1_mode_7_eigval.tex
3.8

%% file: data/ring_Q0_1_mode_8_eigval.tex
21.05

%% file: data/ring_I_mode_1_eigval.tex
0

%% file: data/ring_I_mode_2_eigval.tex
0.59

%% file: data/ring_I_mode_3_eigval.tex
0.59

%% file: data/ring_I_mode_4_eigval.tex
2

%% file: data/ring_I_mode_5_eigval.tex
2

%% file: data/ring_I_mode_6_eigval.tex
3.41

%% file: data/ring_I_mode_7_eigval.tex
3.41

%% file: data/ring_I_mode_8_eigval.tex
4

%% file: data/ring_Q10_mode_1_eigval.tex
0

%% file: data/ring_Q10_mode_2_eigval.tex
0.24

%% file: data/ring_Q10_mode_3_eigval.tex
0.59

%% file: data/ring_Q10_mode_4_eigval.tex
1.31

%% file: data/ring_Q10_mode_5_eigval.tex
2

%% file: data/ring_Q10_mode_6_eigval.tex
2.8

%% file: data/ring_Q10_mode_7_eigval.tex
3.41

%% file: data/ring_Q10_mode_8_eigval.tex
3.85

%% file: clustering_graph.tex
\begin{figure}[t]
  \centering
  \begin{subfigure}{0.46\linewidth}
    \centering
    \includegraphics{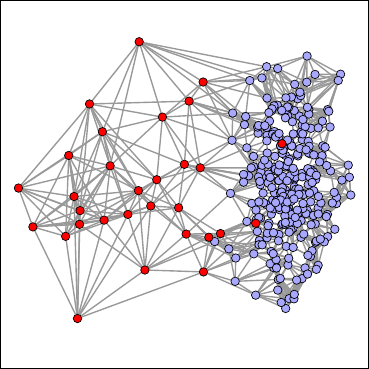}
    \caption{Ground Truth}\label{fig:experiments:clustering:graph:classes}
  \end{subfigure}~
  \begin{subfigure}{0.5\linewidth}
    \centering
    \includegraphics{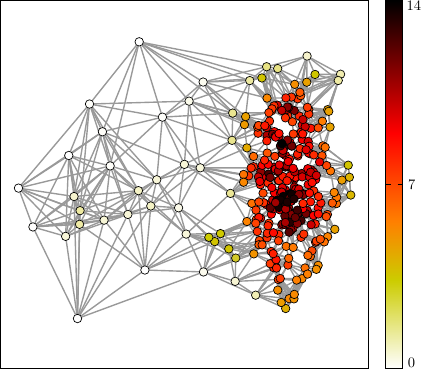}
    \caption{Degree}\label{fig:experiments:clustering:graph:degree}
  \end{subfigure}
  \caption{Two cluster dataset input with a sparse cluster (left) and a dense cluster (right). 
    \subref{fig:experiments:clustering:graph:classes} Ground truth clusters. 
    \subref{fig:experiments:clustering:graph:degree} Vertex degrees from the 10-nearest neighbor graph Gaussian weights.}
  \label{fig:experiments:clustering:graph}
\end{figure}

%% file: clustering_clusters.tex
\begin{figure*}[t]
  \centering
  \begin{subfigure}{0.16\linewidth}
    \centering
    \hfill
    \includegraphics{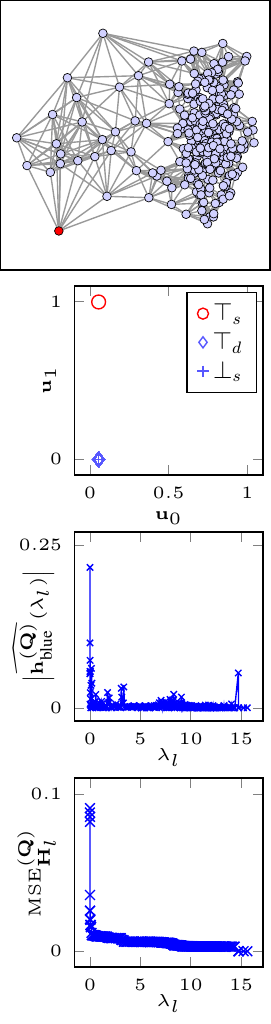}
    \caption{$(\lapl,\eye)$}\label{fig:experiments:clustering:lapl}
  \end{subfigure}
  \begin{subfigure}{0.16\linewidth}
    \centering
    \hfill
    \includegraphics{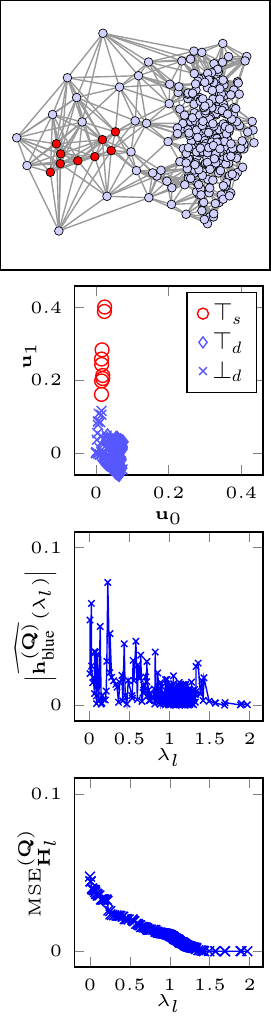}
    \caption{$(\normlapl,\eye)$}\label{fig:experiments:clustering:normlapl}
  \end{subfigure}
  \begin{subfigure}{0.16\linewidth}
    \centering
    \hfill
    \includegraphics{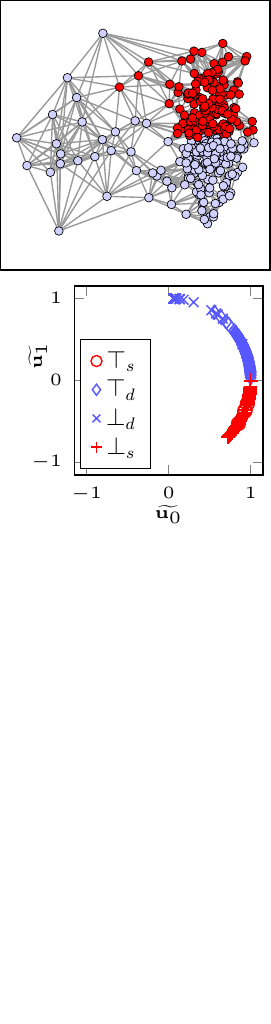}
    \caption{$(\normlapl,\eye)$, feat. norm.}\label{fig:experiments:clustering:normlapl_orig}
  \end{subfigure}
  \begin{subfigure}{0.16\linewidth}
    \centering
    \hfill
    \includegraphics{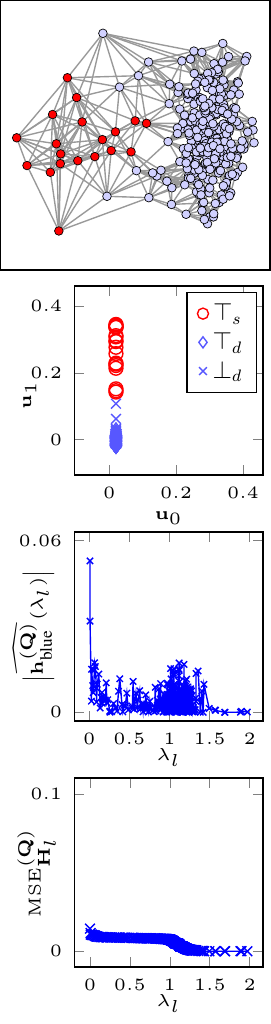}
    \caption{$(\lapl,\degree)$}\label{fig:experiments:clustering:rwlapl}
  \end{subfigure}
  \begin{subfigure}{0.16\linewidth}
    \centering
    \hfill
    \includegraphics{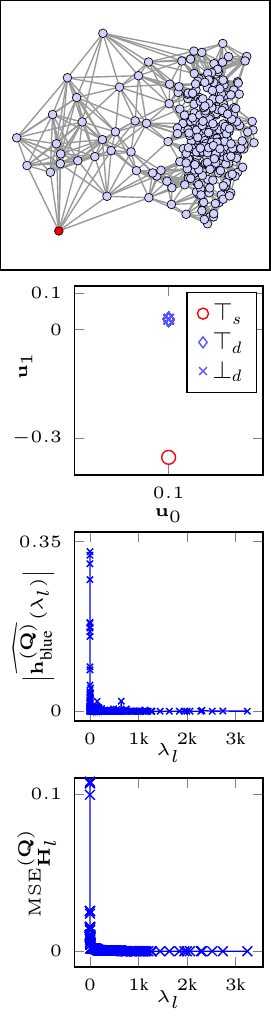}
    \caption{$(\lapl,\Cbf)$}\label{fig:experiments:clustering:voro}
  \end{subfigure}
  \begin{subfigure}{0.16\linewidth}
    \centering
    \includegraphics{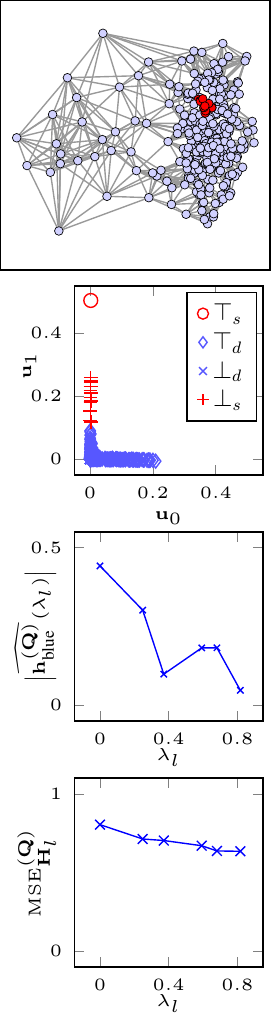}
    \hspace{-0.05cm}\caption{$(\|\xbf-\adjanorm\xbf\|_1,\eye)$}\label{fig:experiments:clustering:a1}
  \end{subfigure}
  \caption{Several spectral clustering results on the graph of \autoref{fig:experiments:clustering:graph}. The smallest cluster 
    is in dark red, and the largest cluster is in light blue. The second row of plots shows the input in the feature space, \ie, 
    to each vertex $i$ is associated a point $\cramped{\bigl[[\ulbf{0}]_i,[\ulbf{1}]_i\bigr]}^T$ in the spectral feature space. 
    The set $\top_s$ (resp. $\top_d$) corresponds to the vertices in the sparse (resp. dense) cluster and correctly clustered. 
    The set $\bot_s$ (resp. $\bot_d$) corresponds to vertices incorrectly clustered in the sparse (resp. dense) cluster. Colors in 
    both rows of figures match for consistency. 
    The third row shows the \textgftaccro{\smoothmat}{\qmatrix} of the corresponding normalized indicator function 
    $\hbf^{\smash{(\qmatrix)}}_{\text{blue}}$ (the first Fourier component is not shown as it is quite large, except for 
    \subref{fig:experiments:clustering:a1}).
    The fourth row shows the \qmatrix-MSE of the ideal low-pass filter $\Hbf_l$ with varying cut-off frequency $\lambda_l$ 
    applied to this normalized indicator function.
    Generalized GFT:
    \subref{fig:experiments:clustering:lapl}: $(\lapl,\eye)$, 
    \subref{fig:experiments:clustering:normlapl} $(\normlapl,\eye)$, \subref{fig:experiments:clustering:normlapl_orig} 
    $(\normlapl,\eye)$ with normalized feature vectors, \subref{fig:experiments:clustering:rwlapl} $(\lapl,\degree)$, 
    \subref{fig:experiments:clustering:voro} $(\lapl,\Cbf)$, and \subref{fig:experiments:clustering:a1} $(\smoothness,\eye)$ 
    with $\smoothness(\xbf)=\|\xbf-\adjanorm\xbf\|_1$.}
  \label{fig:experiments:clustering}
\end{figure*}

%% file: clustering_MSE_I.tex
\begin{figure}[tb]
  \centering
  \hfill
  \begin{subfigure}{0.45\linewidth}
    \centering
    \includegraphics{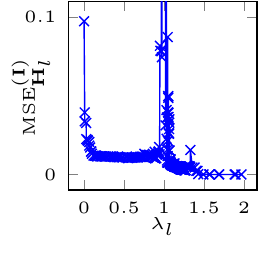}
    \vspace{-0.3cm}
    \caption{$(\lapl,\degree)$}
    \label{fig:experiments:clustering_MSE_I:rwlapl}
  \end{subfigure}
  \hfill
  \begin{subfigure}{0.45\linewidth}
    \centering
    \includegraphics{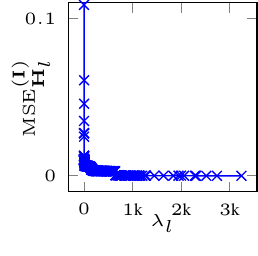}
    \vspace{-0.3cm}
    \caption{$(\lapl,\Cbf)$}
    \label{fig:experiments:clustering_MSE_I:voro}
  \end{subfigure}
  \hfill
  \caption{\eye-MSE of the ideal low-pass filter $\Hbf_l$ applied to the normalized indicator function 
    $\hbf^{\smash{(\qmatrix)}}_{\text{blue}}$ corresponding to \subref{fig:experiments:clustering_MSE_I:rwlapl}
    \autoref{fig:experiments:clustering:rwlapl} and \subref{fig:experiments:clustering_MSE_I:voro} to 
    \autoref{fig:experiments:clustering:voro}.}
  \label{fig:experiments:clustering_MSE_I}
\end{figure}

%% file: sensor_net_figures.tex
\begin{figure*}[tb]
  \centering
  \begin{subfigure}{0.2\linewidth}
    \centering
    \includegraphics{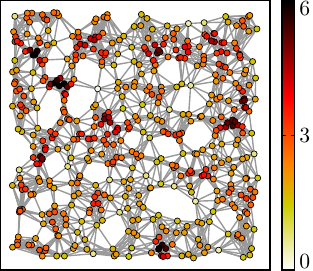}
    \captionsetup{margin={0ex, 3ex}}
    \caption{Degree}
    \label{fig:experiments:sensor_net:energy:unif:degree}
  \end{subfigure}
  \begin{subfigure}{0.2\linewidth}
    \centering
    \includegraphics{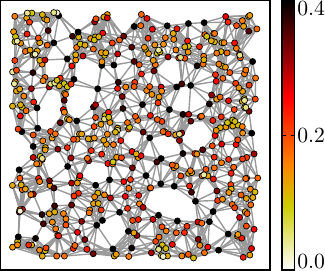}
    \captionsetup{margin={0ex, 3ex}}
    \caption{Voronoi Cell Area}
    \label{fig:experiments:sensor_net:energy:unif:voronoi}
  \end{subfigure}
  \begin{subfigure}{0.29\linewidth}
    \centering
    \includegraphics{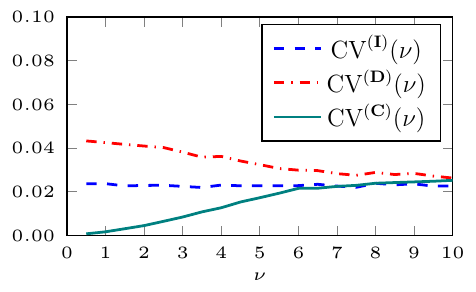}
    \vspace{-0.1cm}
    \captionsetup{margin={3ex, 0ex}}
    \caption{}
    \label{fig:experiments:sensor_net:energy:unif:energy_cv}
  \end{subfigure}
  \begin{subfigure}{0.29\linewidth}
    \centering
    \includegraphics{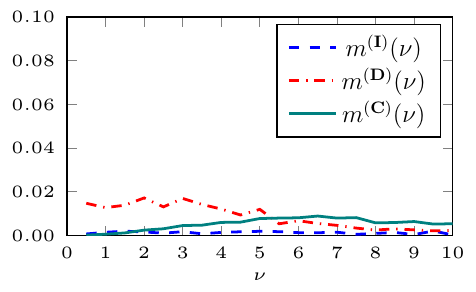}
    \vspace{-0.1cm}
    \captionsetup{margin={3ex, 0ex}}
    \caption{}
    \label{fig:experiments:sensor_net:energy:unif:energy_deviation}
  \end{subfigure}
  \begin{subfigure}{0.2\linewidth}
    \centering
    \includegraphics{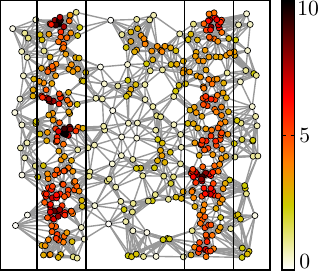}
    \captionsetup{margin={0ex, 3ex}}
    \caption{Degree}
    \label{fig:experiments:sensor_net:energy:non_unif:degree}
  \end{subfigure}
  \begin{subfigure}{0.2\linewidth}
    \centering
    \includegraphics{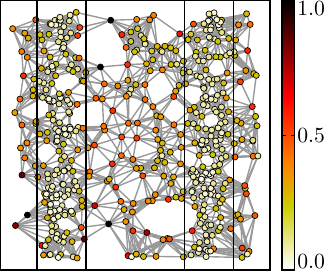}
    \captionsetup{margin={0ex, 3ex}}
    \caption{Voronoi Cell Area}
    \label{fig:experiments:sensor_net:energy:non_unif:voronoi}
  \end{subfigure}
  \begin{subfigure}{0.29\linewidth}
    \centering
    \includegraphics{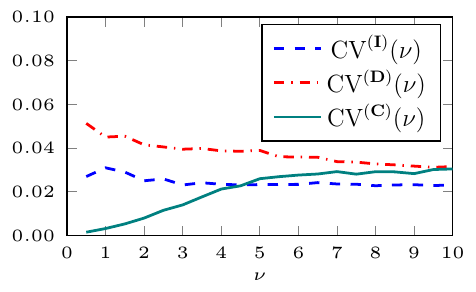}
    \vspace{-0.1cm}
    \captionsetup{margin={3ex, 0ex}}
    \caption{}
    \label{fig:experiments:sensor_net:energy:non_unif:energy_cv}
  \end{subfigure}
  \begin{subfigure}{0.29\linewidth}
    \centering
    \includegraphics{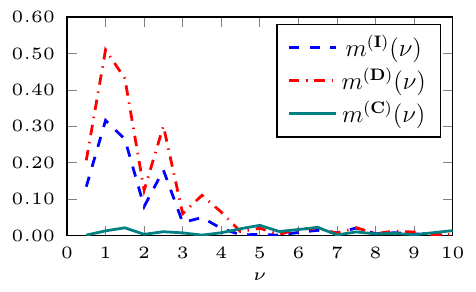}
    \vspace{-0.1cm}
    \captionsetup{margin={3ex, 0ex}}
    \caption{}
    \label{fig:experiments:sensor_net:energy:non_unif:energy_deviation}
  \end{subfigure}
  
  \caption{Study of the graph energy of single cosine continuous signal 
    $\cramped{\widetilde{s_\nu}(x,y;\varphi)=\cos(2\pi\nu x_i+\varphi)}$ sampled uniformly 
    \subref{fig:experiments:sensor_net:energy:unif:degree}-\subref{fig:experiments:sensor_net:energy:unif:energy_deviation} and 
    non-uniformly \subref{fig:experiments:sensor_net:energy:non_unif:degree}%
    -\subref{fig:experiments:sensor_net:energy:non_unif:energy_deviation} as 
    $\cramped{s_\nu(i;\varphi)=\widetilde{s_\nu}(x_i,y_i;\varphi)}$.
    Example sampling with 500 samples (vertices) showing vertex degree 
    (\subref{fig:experiments:sensor_net:energy:unif:degree} and \subref{fig:experiments:sensor_net:energy:non_unif:degree}), and 
    Voronoi cell area (\subref{fig:experiments:sensor_net:energy:unif:voronoi} and 
    \subref{fig:experiments:sensor_net:energy:non_unif:voronoi}) with colors. 
    \subref{fig:experiments:sensor_net:energy:unif:energy_cv}%
    -\subref{fig:experiments:sensor_net:energy:unif:energy_deviation}
    and
    \subref{fig:experiments:sensor_net:energy:non_unif:energy_cv}%
    -\subref{fig:experiments:sensor_net:energy:non_unif:energy_deviation} 
    1000 samples (vertices), with results averaged over 500 sampling (graph) realizations. 
    \subref{fig:experiments:sensor_net:energy:unif:energy_cv} and
    \subref{fig:experiments:sensor_net:energy:non_unif:energy_cv} 
    coefficient of variation $\CV^{(\qmatrix)}(\nu)$ of the signal energy depending on the continuous frequency $\nu$. 
    \subref{fig:experiments:sensor_net:energy:unif:energy_deviation} and
    \subref{fig:experiments:sensor_net:energy:non_unif:energy_deviation} 
    maximum absolute deviation $m^{(\qmatrix)}(\nu)$ of the normalized mean signal energy depending on the continuous frequency 
    $\nu$.}
  \label{fig:experiments:sensor_net:energy}
\end{figure*}

%% file: clustering_lowpass.tex
\begin{figure*}[t]
  \centering
  \setlength\tabcolsep{5pt}
  \def\arraystretch{3}
  \begin{tabular}{rccccc}
    & $(\lapl,\eye)$ & $(\normlapl,\eye)$ & $(\lapl,\degree)$ & $(\lapl,\Cbf)$ & $(\|\xbf-\adjanorm\xbf\|_1,\eye)$\hspace{0.4cm} \\
    
    \rotatebox{90}{\hspace{1cm}$\smash{l=1}$}
    &
    \includegraphics{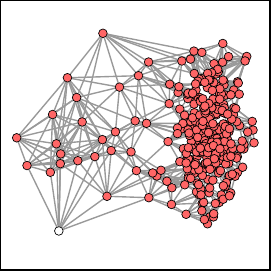}
    &
    \includegraphics{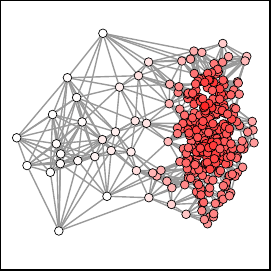}
    &
    \includegraphics{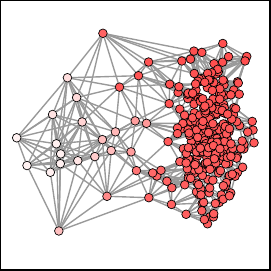}
    &
    \includegraphics{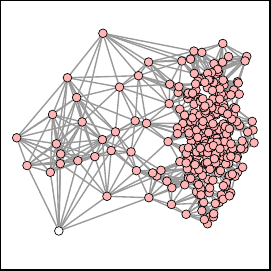}
    &
    \includegraphics{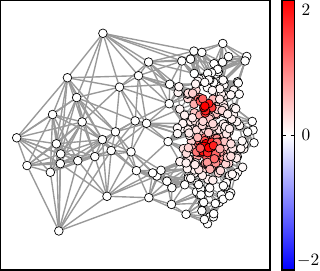}
    \\
  
    \rotatebox{90}{\hspace{1cm}$\smash{l=5}$}
    &
    \includegraphics{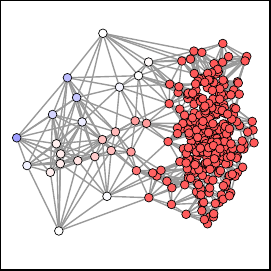}
    &
    \includegraphics{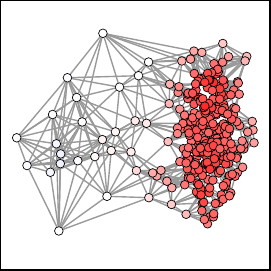}
    &
    \includegraphics{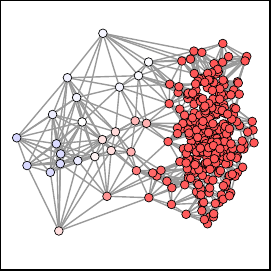}
    &
    \includegraphics{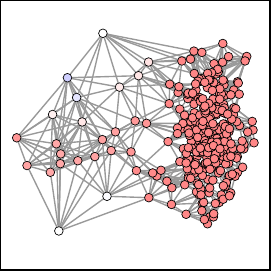}
    &
    \includegraphics{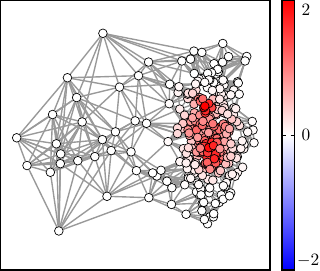}
    \\
  
    \rotatebox{90}{\hspace{1cm}$\smash{l=50}$}
    &
    \includegraphics{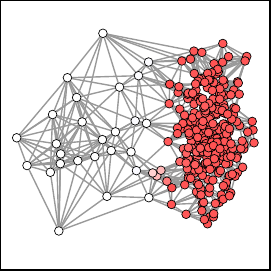}
    &
    \includegraphics{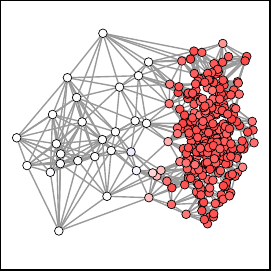}
    &
    \includegraphics{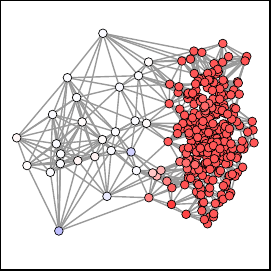}
    &
    \includegraphics{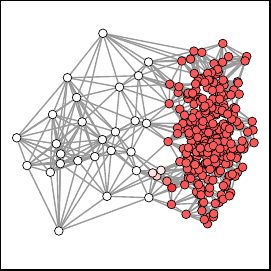}
    &
  \end{tabular}
  \caption{Output of the ideal low pass filter $\Hbf_l$ applied to the normalized indicator function 
    $\hbf^{\smash{(\qmatrix)}}_{\text{blue}}$. $\Hbf_l$ keeps only the first $l+1$ spectral components (\ie, up to the graph
    frequency $\lambda_l$). Note that $\Hbf_l$ depends on the \textgftaccro{\smoothness}{\qmatrix}.}
  \label{fig:experiments:clustering_indicator_low_pass}
\end{figure*}